\documentclass[sigplan,screen]{acmart}
\copyrightyear{2024}
\acmYear{2024}
\setcopyright{acmlicensed}\acmConference[ASPLOS '24]{29th ACM International Conference on Architectural Support for Programming Languages and Operating Systems, Volume 2}{April 27-May 1, 2024}{La Jolla, CA, USA} \acmBooktitle{29th ACM International Conference on Architectural Support for Programming Languages and Operating Systems, Volume 2 (ASPLOS '24), April 27-May 1, 2024, La Jolla, CA, USA}
\acmDOI{10.1145/3620665.3640362} \acmISBN{979-8-4007-0385-0/24/04}

\usepackage[]{hyperref}

\usepackage{algorithmic}
\usepackage{graphicx}
\usepackage{braket}
\usepackage{textcomp}
\usepackage{xcolor}
\usepackage{svg}
\usepackage{subfig}
\usepackage{array}
\usepackage{makecell}
\usepackage{enumitem}

\usepackage[normalem]{ulem}

\author{Sophia Fuhui Lin}
\email{sophialin1@uchicago.edu}
\affiliation{%
  \institution{University of Chicago}
  \country{USA}
}
\author{Joshua Viszlai}
\email{viszlai@uchicago.edu}
\affiliation{%
  \institution{University of Chicago}
  \country{USA}
}

\author{Kaitlin N. Smith}
\email{kaitlin.smith@infleqtion.com}
\affiliation{%
  \institution{Super.tech, a software division of Infleqtion}
  \country{USA}
}

\author{Gokul Subramanian Ravi}
\email{gravi@uchicago.edu}
\affiliation{%
  \institution{University of Chicago}
  \country{USA}
}

\author{Charles Yuan}
\email{chenhuiy@csail.mit.edu}
\affiliation{%
  \institution{MIT CSAIL}
  \country{USA}
}

\author{Frederic T. Chong}
\email{chong@cs.uchicago.edu}
\affiliation{%
  \institution{University of Chicago}
  \country{USA}
}

\author{Benjamin J. Brown}
\email{benjamin.brown@ibm.com}
\affiliation{%
  \institution{IBM Quantum, T. J. Watson Research Center}
  \country{USA}
}
\affiliation{%
  \institution{IBM Denmark}
  \country{Denmark}
}

\renewcommand{\shortauthors}{Lin et al.}
\begin{document}

\title{Codesign of quantum error-correcting codes and modular chiplets in the presence of defects}

\begin{abstract}
Fabrication errors pose a significant challenge in scaling up solid-state quantum devices to the sizes required for fault-tolerant (FT) quantum applications. To mitigate the resource overhead caused by fabrication errors, we combine two approaches: (1) leveraging the flexibility of a modular architecture, (2) adapting the procedure of quantum error correction (QEC) to account for fabrication defects.

We simulate the surface code adapted to defective qubit arrays to find metrics that characterize how defects affect fidelity. We then use our simulations to determine the impact of defects on the resource overhead of realizing a fault-tolerant quantum computer on a chiplet-based modular architecture.
Our QEC simulation adapts the syndrome readout circuit for the surface code to account for an arbitrary distribution of defects.
Our simulations show that our strategy for dealing with fabrication defects demonstrates an exponential suppression of logical failure, where error rates of non-defective physical qubits are $\sim 0.1\%$ for a circuit-based noise model. This is a typical regime on which we imagine running the defect-free surface code.
We use our numerical results to establish post-selection criteria for assembling a device with defective chiplets. Using our criteria, we then evaluate the resource overhead in terms of the average number of physical qubits fabricated for a logical qubit to obtain a target logical error rate.
We find that an optimal choice of chiplet size, based on the defect rate and target performance, is essential to limiting any additional error correction overhead due to defects. When the optimal chiplet size is chosen, at a defect rate of $1\%$ the resource overhead can be reduced to below 3X and 6X respectively for the two defect models we use, for a wide range of target performance. Without tolerance to defects, the overhead grows exponentially as we increase the number of physical qubits in each logical qubit to achieve better performance, and also grows faster with an increase in the defect rate. When the target code distance is 27, the resource overhead of the defect-intolerant, modular approach is 45X and more than $10^5$X higher than the super-stabilizer approach, respectively, at a defect rate of $0.1\%$ and $0.3\%$.
We also determine cutoff fidelity values that help identify whether a qubit should be disabled or kept as part of the QEC code.
\end{abstract}
\maketitle

\section{Introduction}
Fault-tolerant (FT) quantum computers will enable the implementation of large-scale quantum algorithms such as search~\cite{shor1999polynomial} and factoring~\cite{grover1996fast}. These machines are designed to protect quantum information by encoding logical qubits in quantum error-correcting codes that consist of a large number of interacting physical qubits.

Achieving fault tolerance requires a practical way to manufacture physical qubits.
One leading approach is to manufacture a large planar array~\cite{fowler2012surface} of qubits in a solid-state system such as superconducting~\cite{Krinner2022, sundaresan2022matching,Google2023} or spin qubits~\cite{Froning2021, Hendrickx2021}. In these hardware architectures, Ref.~\cite{o2017quantum} estimates that over a million high-fidelity qubits will be required for FT quantum applications. However, scaling a quantum device to this magnitude presents significant challenges.


The fabrication of large-scale solid-state devices encounters additional complexities due to fabrication errors. There are many steps in processing where a slight deviation from the target specification occurs due to process imprecision or stochastically appearing impurities or imperfection\cite{hertzberg2021laser, kreikebaum2020improving}. As a result, variation in the quality of qubits, as well as the links along which qubits interact, is inevitable. On quantum devices, physical qubits will exhibit inhomogeneous characteristics, including variations in gate success, measurement fidelity, and coherence, and will likely contain \textit{faulty (defective) qubits} --- qubits with severely limited functionality. With today's technology, \cite{kate_chiplet} estimated that $\sim 2\%$ of the qubits on a transmon device would be faulty.
To effectively scale up solid-state quantum devices in the presence of fabrication errors, we combine two approaches: (1) Leveraging the flexibility of a modular architecture, which offers greater adaptability compared to a monolithic architecture by enabling the post-selection of individual chiplets prior to their integration into a larger device. (2) Adapting QEC codes to defective qubit arrays, which enables the utilization of some defective chiplets. The combination of these two strategies helps mitigate the additional resource overhead caused by fabrication errors, but it requires a thorough understanding of the performance of an adapted surface code, and how the overhead is affected by design choices. Importantly, establishing an informed post-selection criterion that accounts for how each chiplet is affected by its defects requires identifying performance indicators specific to the adapted surface code. This also helps us understand how the logical fidelity of an adapted surface code compares with that of a standard surface code.

\begin{figure}
\centering
\subfloat[]{\includegraphics[width=1.0in]{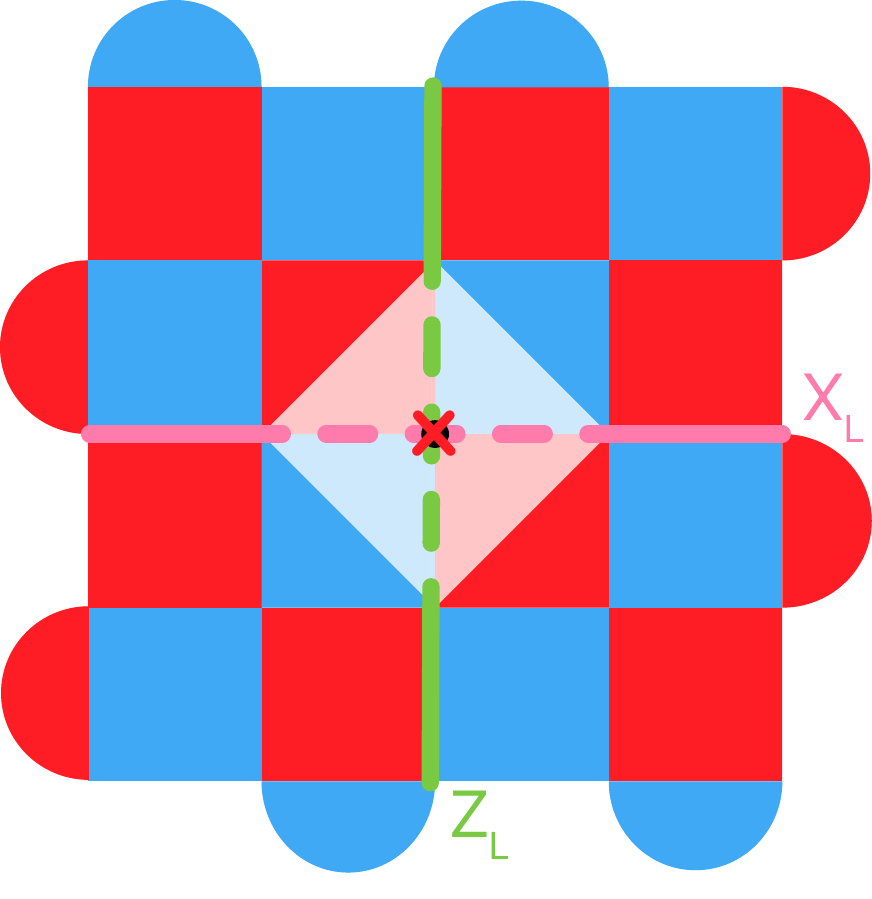}}\qquad
\subfloat[]{\includegraphics[width=1.0in]{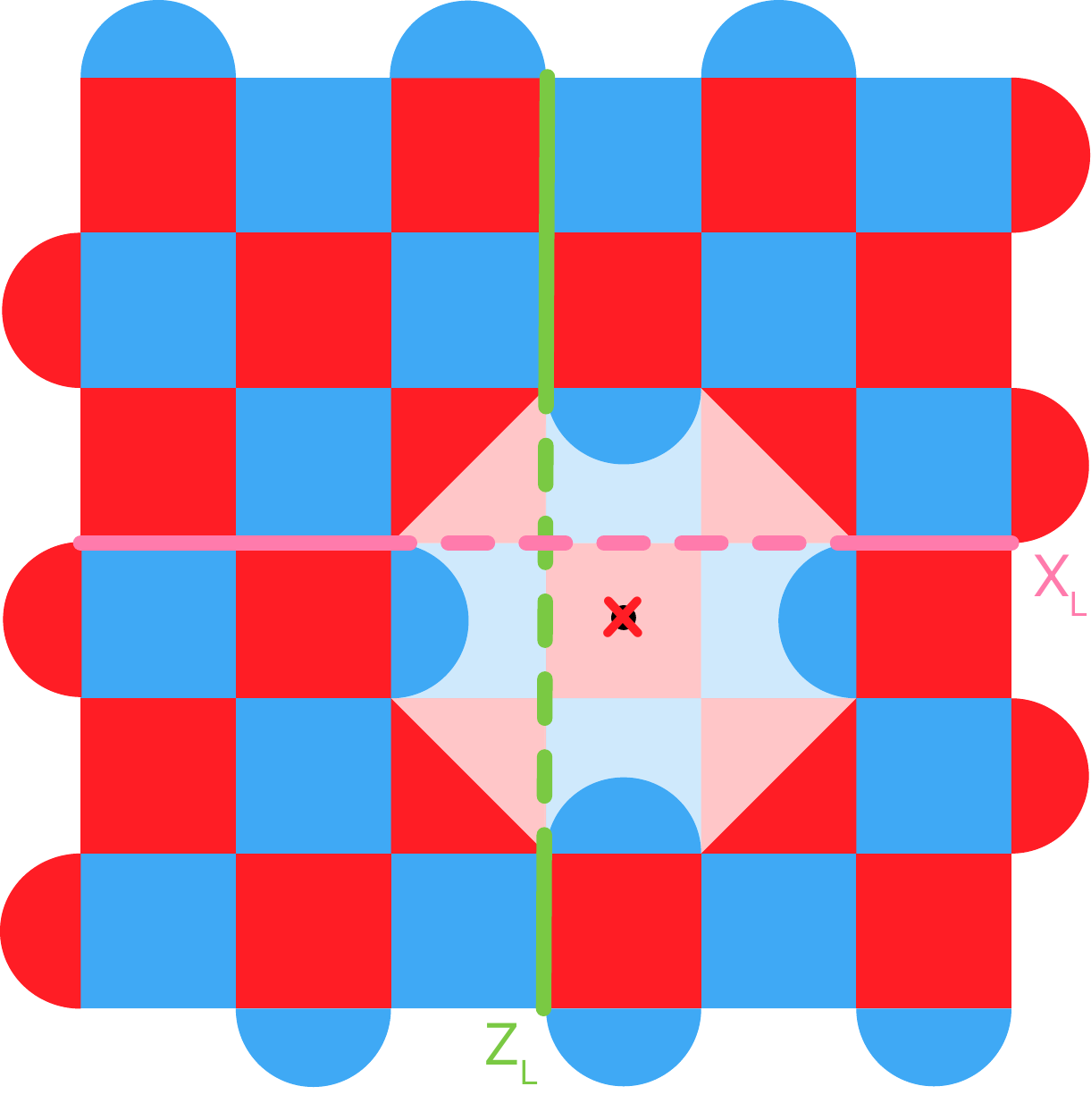}}\qquad
\subfloat[]{\includegraphics[width=1.0in]{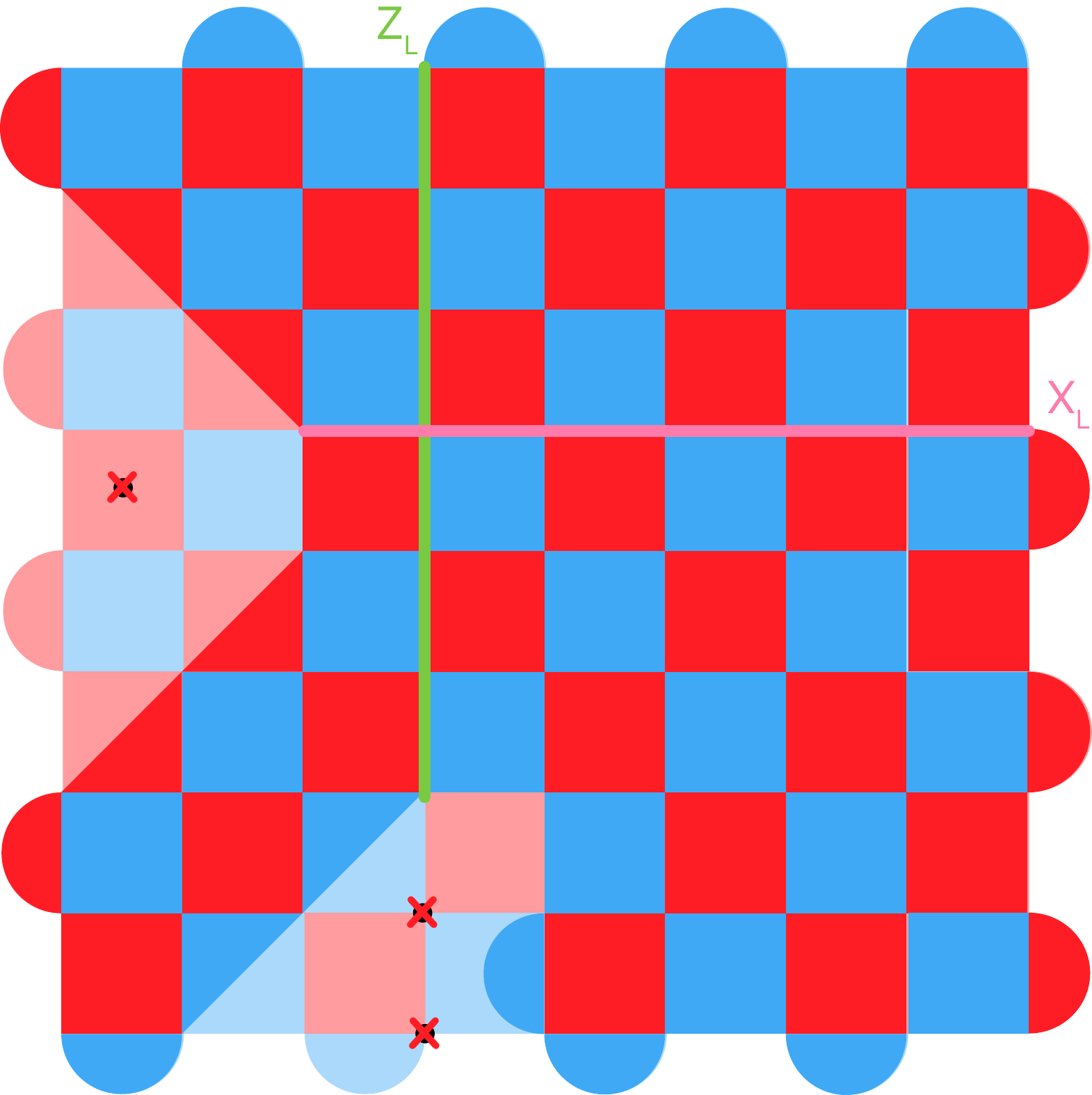}}\qquad
\subfloat[]{\includegraphics[width=1.0in]{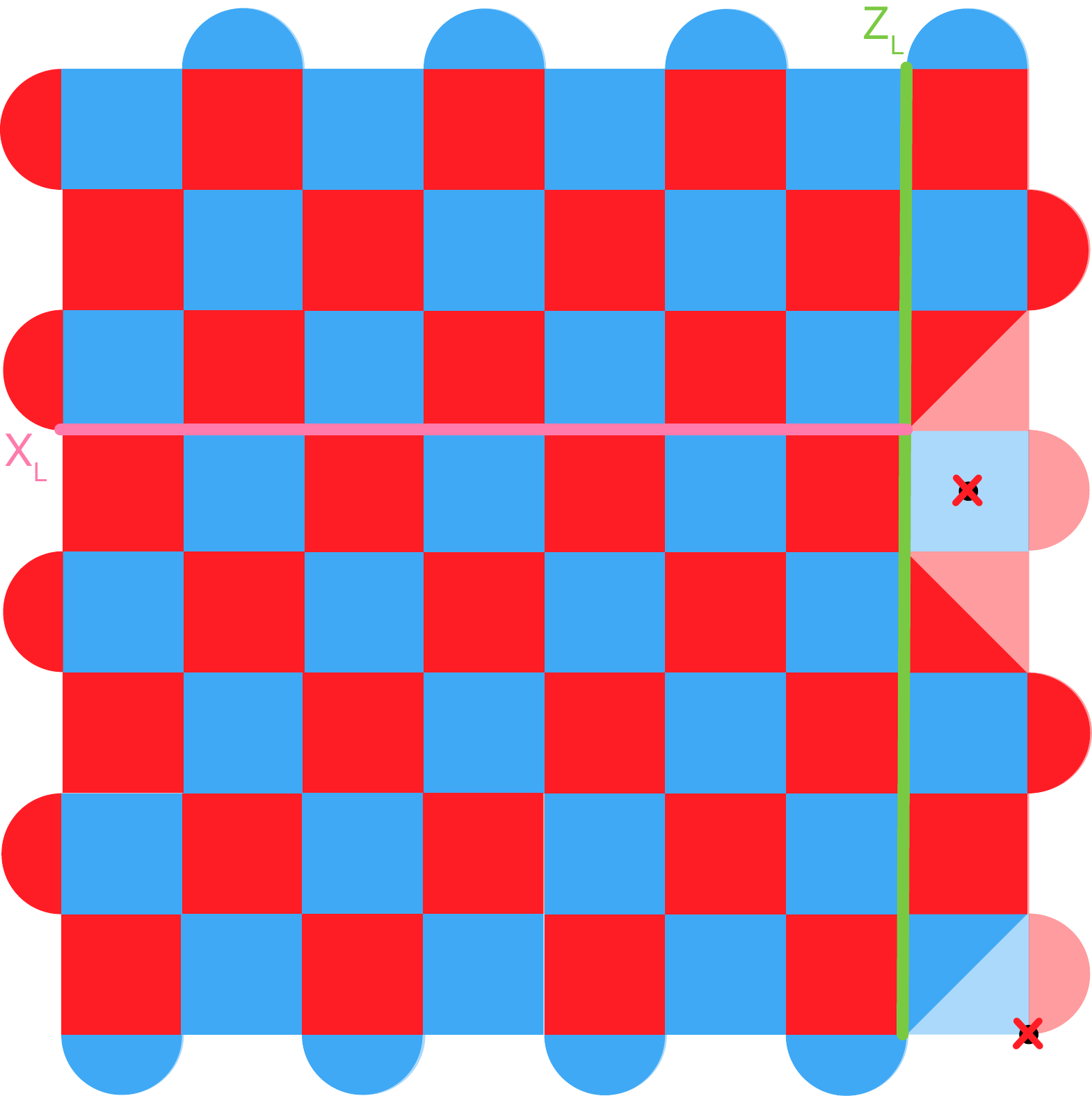}}
\caption{Examples of super-stabilizers and boundary deformations. The faulty qubits are marked with a red `x', and the excluded part in a patch is represented by lighter color. (a) One broken data qubit in the interior, handled by a super-stabilizer. (b) One broken syndrome qubit in the interior, handled by larger super-stabilizers. (c)(d) Broken qubits near the boundary require boundary deformations.}
\label{fig:super-stabilizer}
\end{figure}

An adapted surface code can be implemented on a defective chiplet by employing ``super-stabilizer measurements,'' which are capable of detecting errors in proximity to defective areas, as outlined in previous studies~\cite{Stace2009,Stace2010,dan2017, shell_theory}. We illustrate the construction of super-stabilizers in Fig.~\ref{fig:super-stabilizer}. Conveniently, these super-stabilizers can be measured in a timely manner using the operational qubits adjacent to the defective region~\cite{Stace2009,Stace2010,dan2017}. It has been shown, in theory, that this approach can scale on a large lattice~\cite{shell_theory}. However, there is a noticeable gap in existing research regarding the comparative performance and resource overhead of such adapted codes relative to the standard, defect-free surface code. To address this, we conduct numerical simulations. And to facilitate our numerical simulations, we develop an automated method to map surface code to defective grids by deforming boundaries and forming super-stabilizers. Notably, our automated method can define a surface code for an arbitrary configuration of defects.

Using our numerical results, we identify two key figures of merit that characterize the fidelity of a surface code on a defective chiplet. The first is the distance on the defective patch $d$, the least number of physical errors that can lead to a logical failure. In the regime of low error rate per gate, $p$, we find that to leading order, the logical failure rate decays like $O(p^{d/2})$, just as for defect-free chiplets. Remarkably, we identify this scaling when physical errors occur at a rate $p\sim 10^{-3}$, the regime where we expect a defect-free chip to be operable. This implies that the defective chips are functionally similar to the defect-free ones with the same $d$, except that they cost more physical qubits. Aside from this scaling, we identify variation in surface codes with equivalent code distances. We find a second figure of merit that differentiates among these codes. Specifically, we find that the logical failure rate will scale with the number of different ways that a logical failure can occur with $d$ physical errors. Both of these figures of merit can be efficiently computed after the surface code is adapted to a defective grid. 

These two indicators enable us to rapidly assess the quality of individual defective chiplets. This is necessary for establishing a post-selection criterion for the modular chiplet architecture, and also for resource overhead evaluation. 
Our numerical results demonstrate that our post-selection criterion is more effective than the natural strategy to select the chiplets with the fewest defects. 

We quantify the resource overhead by the total number of fabricated physical qubits per logical qubit, including the qubits on the chiplets that are not selected.
In Fig.\ref{subfig:yield_fixed_target9_cost},\ref{subfig:yield_tunable_target9_cost}, and \ref{subfig:yield_fixed_target17_cost}, we show the factor of resource overhead relative to the ideal no-defect case. Our results show that it is important to select the chiplet size based on the fabrication error rate, in order to achieve a balance between having a high yield (proportion of chiplets that meet the standard) and using a small number of physical qubits per patch of code. When the optimal chiplet size is chosen based on the results, the resource overhead is below 3X and 6X respectively for the two defect models we study (defective links only v.s. same rate of defect on links and qubits), when the defect rate is below $1\%$. This holds for a wide range of performance targets. Without the ability to tolerate defects, the resource overhead grows exponentially with the number of physical qubits in a logical qubit (which increases for higher performance targets) and also grows faster with the defect rate.

We also evaluate the sensitivity of the yield to two design choices: what boundary constraint is imposed on each patch, and whether the chiplet design allows the data and syndrome qubits to be swapped by rotating a chiplet. When the architecture allows swapping the assignment of syndrome and data qubits for individual logical qubits, for example by rotating a chiplet, the yield can be improved. Although in this paper we mainly evaluate each patch by its capacity to store a logical qubit, a boundary constraint can ensure the quality of multi-qubit logical gates.

Finally, we identify cutoff fidelity values for determining whether a qubit should be treated as faulty or kept in the code. This is important in the practical setting because the impact of fabrication errors varies and there's not always a clear line between faulty and good qubits.

\textbf{Summary of contributions and results:}
\begin{itemize}[noitemsep,topsep=-\parskip]
\item We develop an automated method that adapts the rotated surface code to a grid with an arbitrary distribution of defects using super-stabilizers. Our automated method produces a simulation of active error correction that is implemented on Stim~\cite{gidney2021stim}. Our code and a demo notebook is available at \url{https://github.com/SophLin/superstabilizer_demo}.
\item Using our numerical simulations, we identify effective indicators for assessing the fidelity of a surface code adapted to defective chiplets. 
\item We present the first evaluation of the impact of fabrication defects on the resource overhead of quantum error correction. Our focus is on the modular chiplet architecture where one logical qubit is allocated on each chiplet. We quantify the resource overhead as the average number of physical qubits fabricated for a logical qubit, and evaluate its sensitivity to system parameters. We show that the post-selection of chiplets and the ability to use defective chips are both critical for reducing the extra overhead caused by fabrication errors. 
\item We identify cutoff fidelity values for determining whether a qubit with worse performance than its neighbors should be treated as faulty or kept in the code.
\end{itemize}
\section{Background}

\subsection{Fabrication errors and variations on transmon-based quantum devices}
\label{sec:transmon_bg}


In this section, we discuss some of the sources of fabrication errors and variation for transmon qubits.  Although our discussion is not exhaustive, it is meant to give some intuition for why current chips see a 2\% defect rate.

Imprecision in the fabrication of Josephson Junctions (JJs) is a source of varied qubit performance. A JJ, two superconductors separated by a thin metal-oxide insulator, is the heart of a transmon qubit~\cite{gambetta2017building}. JJs have incredibly small feature sizes that are hundreds of nanometers in scale~\cite{hertzberg2021laser}, smaller than the wavelengths used during optical lithography. Thus, slight imperfections that appear in JJ positioning, component dimension, or surrounding layers influence operational characteristics of the transmon~\cite{kreikebaum2020improving}.

On fixed-frequency transmons with fixed couplers, one of the  most commonly used superconducting qubits, frequency collision is a dominant type of fabrication error. Fabrication variation can deviate a qubit's frequency from the ideal frequency, resulting in spectral overlaps that cause frequency collisions. This variation is stochastic, causing the resulting frequency profile of each chip to be unique. 
Another type of unintended defect that frequently and stochastically appears across a quantum chip during processing is a two-level system (TLS)~\cite{muller2019towards}. A TLS is caused by impurities inside materials or irregularities within atomic crystalline lattice structures appearing unexpectedly in oxide layers or on the surface of the chip. Because of the layered approach associated with transmon processing, there are many opportunities for TLSs to appear during fabrication.

\subsection{Surface code}
\label{bg_qec}
\begin{figure}
  \centering
  \includegraphics[width=0.12\textwidth]{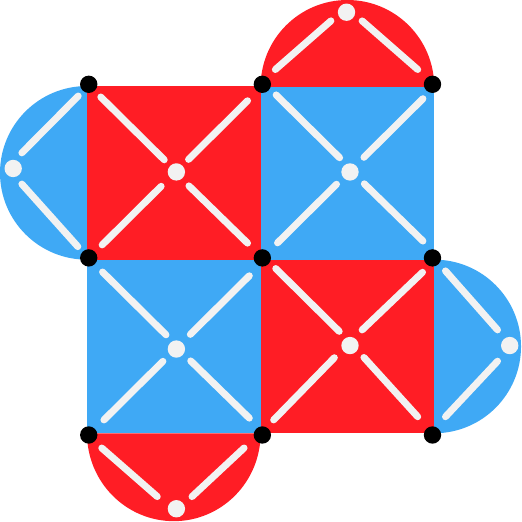}
  \caption{Rotated surface code with d=3. The black dots represent the data qubits, and each red (blue) face represents an X (Z) stabilizer. Each stabilizer requires an ancilla qubit.}
  \label{fig:surface_code_background}
\end{figure}

The surface code~\cite{Dennis02,fowler2012surface,Kitaev03} is one of the most practical quantum error-correcting codes for physical realization due to its implementation using a two-dimensional nearest-neighbor qubit layout, and its high tolerance to noise.
It can perform a universal set of logic gates while maintaining its local planar layout. One can use lattice surgery to perform entangling operations~\cite{Horsman_2012, Brown17, Litinski2019} and magic state distillation~\cite{Bravyi05, Gidney2019} to perform non-Clifford gates. For most of this work, we will concentrate on the performance of the surface code storing a logical qubit over time.
In this paper, we use the rotated planar surface code due to its low qubit overhead \cite{Bombin06, fowler2018low, beverland2019role}. We define the surface code with code distance $d$ on a $d\times d$ grid of data qubits. The errors on the data qubits are detected using $d^2-1$ measurement qubits, otherwise known as ancilla qubits or syndrome qubits. More specifically, a measurement qubit is placed on either a red or a blue face of the grid of data qubits, as in Fig.~\ref{fig:surface_code_background}. Note also that the faces at the lattice boundary each touch two data qubits.

We measure stabilizers to detect errors that qubits experience. Stabilizers are measured repeatedly in cycles to determine the locations of errors that occur over time.
In each cycle, each measurement qubit is used to measure either a Pauli-X stabilizer,  $\hat{X}_a\hat{X}_b\hat{X}_c\hat{X}_d$, or a Pauli-Z stabilizer $\hat{Z}_a\hat{Z}_b\hat{Z}_c\hat{Z}_d$, depending on the color of the lattice face on which the measurement qubit lies. The measurement qubits on the boundary only acts on two data qubits, and therefore measure stabilizers of the type $\hat{X}_a\hat{X}_b$ or $\hat{Z}_a\hat{Z}_b$. Stabilizers are measured with circuits that are detailed in, e.g., Ref.~\cite{Tomita2014}.

In practice, to deal with the errors that occur on data qubits, as well as errors that occur on measurement qubits that may cause stabilizer circuits to give unreliable outcomes, we compare a stabilizer reading at cycle $t$ to the reading of the same stabilizer made at cycle $t-1$. 
A difference in reading gives rise to an error detection event.

By performing multiple cycles, we obtain a history of detection events over time that we call the error syndrome. In general, we can regard errors as string-like objects in this error syndrome, where detection events occur at the end-points of these strings. See e.g. Refs.~\cite{Brown2023, Dennis02, fowler2012surface, Wang03} for details. Using the error syndrome, we can obtain a correction to recover encoded information using minimum-weight perfect-matching algorithm~\cite{Brown2023, Dennis02, fowler2012surface, higgott2023sparse, Wang03}, where we deal with detection events due to Pauli-X stabilizers and Pauli-Z stabilizers separately. We concentrate on only Pauli-Z stabilizers throughout this work, but note that an equivalent analysis will hold for the alternative stabilizers.

A logical error is introduced into the surface code when at least $d/2$ errors occur along a non-trivial path over the surface code error syndrome history~\cite{beverland2019role, Dennis02, fowler2012surface,Wang03}. If we assume that an individual error occurs with probability $O(p)$, then in the limit that $p$ is small, we can fit the logical error rate to the ansatz \begin{equation}
LER = \beta (Np)^{\alpha d}, \label{Eqn:FigureOfMerit}
\end{equation}
where $N$ and $\alpha \leq 1/2$ are constants to be determined~\cite{beverland2019role, Bravyi13simulation, Dennis02, fowler2012surface, Wang03, Watson_2014}. 









\section{Deforming boundary and forming super-stabilizers}\label{sec:superstabilizer}

Faulty qubits and links can be particularly harmful to the implementation of topological QEC codes if they are not handled correctly. This is because qubit errors in the vicinity of fabrication defects may not be detected. Given a finite density of fabrication defects, this will inhibit the decay of logical failure rate as we increase code distance, if we do not deal with these defects correctly.

\begin{figure}
  \centering
  \includegraphics[width=0.4\textwidth]{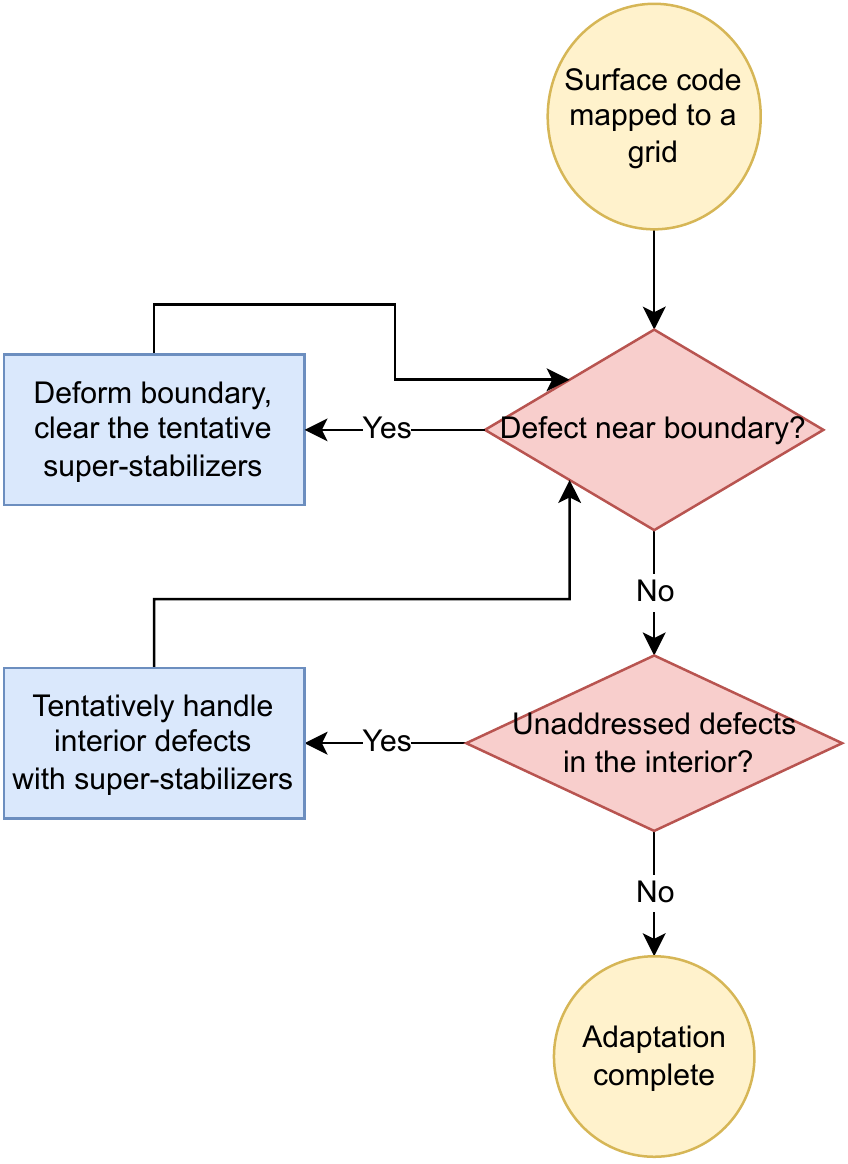}
  \caption{Algorithm for mapping rotated surface code to a defective grid by deforming boundaries and forming super-stabilizers.}
  \label{fig:flowchart}
\end{figure}

Building upon the theory in Ref.~\cite{shell_theory}, we develop and implement an automated method that adjusts a surface code for arbitrary defect distributions. Our algorithm includes deforming boundaries of a code and forming super-stabilizers in the interior. A flowchart is shown in Fig. \ref{fig:flowchart}. Our code takes the chiplet size $l$ and a list of defects as input, then adapts a surface code to the defective grid and generates a stabilizer measurement circuit compatible with the Stim\cite{gidney2021stim} simulator.

We learn the occurrence of errors close to fabrication defects by forming super-stabilizers around them. The values of these super-stabilizers can be inferred from local measurements around the defects~\cite{Stace2009, Stace2010, dan2017}. Furthermore, we repeat the local measurements used to obtain the super-stabilizer for an elongated time, over a timescale that is commensurate with the size of the super-stabilizer that are defective~\cite{shell_theory}. A measurement schedule with this feature is adopted in Ref.~\cite{shell_theory} to demonstrate the procedure helps achieve a threshold --- when the physical errors on the non-faulty qubits is lower than the threshold, one can expect the quality of the logical qubit to improve as it is encoded with more physical qubits. 

As an example, we show how we can construct a weight-6 X and Z super-stabilizer around a single faulty qubit in the interior of the code in Fig. \ref{fig:super-stabilizer}a). The value of the X (Z) super-stabilizer can be computed from the direct product of the two weight-3 broken X (Z) stabilizers used as gauge operators. The X and Z gauges anti-commute, so they cannot be measured in the same cycle. Instead, they are measured in alternate cycles.

Another important example is the case where a measurement qubit in the interior of the qubit grid is faulty (Fig. \ref{fig:super-stabilizer}b). All of its neighboring data qubits are disabled and larger super-stabilizers (each consisting of 4 gauge operators) are formed. Around a larger defect cluster like this, instead of measuring the X and Z gauges in alternate cycles (XZ...), we repeat one type of measurement several times before switching to the other (e.g. XXZZ...) following~\cite{shell_numerics, shell_theory}. The number of repetitions should scale with the size of the cluster~\cite{shell_numerics, shell_theory}; here we set the number of repetitions equal to the diameter of the defect cluster.

Note that when forming super-stabilizers in the interior, we not only need to disable the data qubits connected to defective syndrome qubits, but also need to disable syndrome qubits due to defective data qubits in some cases. It is obvious that a syndrome qubit connected to no more than one active data qubit needs to be disabled. When a syndrome qubit is connected to two active data qubits but the three qubits are on the same diagonal line, it also needs to be disabled.

If a faulty qubit is too close to the boundary to be surrounded by gauge operators, it cannot be handled by super-stabilizers. In this case the boundary of the patch needs to be deformed to exclude the faulty qubit. Although \cite{dan2017} addressed the boundary deformation for a surface code with a different lattice geometry, the rotated surface code has more complicated boundaries and we develop a new algorithm for handling the defects near boundary.

To illustrate a boundary deformation, four examples are shown in Fig. \ref{fig:super-stabilizer} (c) and (d). On the right side of (d) is a faulty syndrome qubit near a boundary of different color. Two data qubits are disabled along with it, because they are no longer included in any Z stabilizer. Then the X syndrome qubit on its right is also disabled, because it no longer has any data qubits to measure. If any of these three qubits were the faulty one, the same boundary deformation would apply. If a faulty syndrome qubit is near a boundary of the same color, as is the case on the left of (c), more qubits need to be excluded from the patch to ensure that all stabilizers on the boundary are of the same color. In particular, the neighboring syndrome qubits of different type than the boundary need to be excluded from the patch. If a data or syndrome qubit at a corner is faulty, then only one other qubit needs to be excluded (lower right of (d)). If any faulty qubit is too close to the new boundary, it must be excluded too. To the lower left of (c), such an example is shown. The faulty data qubit on the original boundary leads to an excluded region that is similar to the one on the right edge of (d). Then since a data qubit on the new boundary is faulty, the lower boundary is further deformed. Note that the second faulty qubit in this region was part of three stabilizers that remained active after the first deformation, but only the one of different color than the lower boundary is excluded in the second deformation.

The code distance $d$ is the length of the shortest undetectable error chain on a patch of QEC code, and is equivalent to the length of the shortest X or Z logical operator. As we will show in Sec. \ref{sec:main_results}, it not only characterizes the defect-free codes, but also serves as a primary indicator for the fidelity of defective patches. On a $l\times l$ patch, we have $d=l$ only if there is no defect; a defective patch has $d < l$. In Fig. \ref{fig:super-stabilizer} (a), $l=5$ and $d=4$ along both directions. In (b), we have $l=7$ and $d=5$. In (c) and (d), $l=9$. The code distance is 7 in (c). In (d), the distance is different along each direction: $d=9$ vertically and $d=8$ horizontally. 
\section{Building a device with defective qubits}\label{sec:main_results}

In this section, we move on to the setting where the goal is to build a large FT device with an array of rotated surface code patches. We first propose a modular architecture and discuss design choices, then identify a post-selection criterion for evaluating the quality of defective chiplets.

For the simulation, we use two models of fabrication errors: one with links set to be faulty at random, and one with links and qubits both set to be faulty at the same probability. The first one models fixed-frequency transmon qubits with fixed couplers, where frequency collision is the dominant type of fabrication error. The latter models tunable transmon qubits, where links are as intricate as qubits. When using the super-stabilizers, a faulty link can be handled by disabling either of the two qubits that it connects. Faulty syndrome qubits lead to greater damage as explained in the last section, so we choose to disable the data qubit connected to a faulty link unless the syndrome qubit on the other end is already disabled.

We use a circuit-level noise model for the physical errors on the non-defective qubits, where the two-qubit gate error is $p$, the one-qubit gate error is $0.8p$, and the readout error is $\frac{8}{15}p$. We use standard measurement circuits for the syndromes~\cite{Tomita2014}.

\subsection{A modular architecture for rotated surface code}\label{sec:describe_arch}
\begin{figure}
  \centering
  \includegraphics[width=0.4\textwidth]{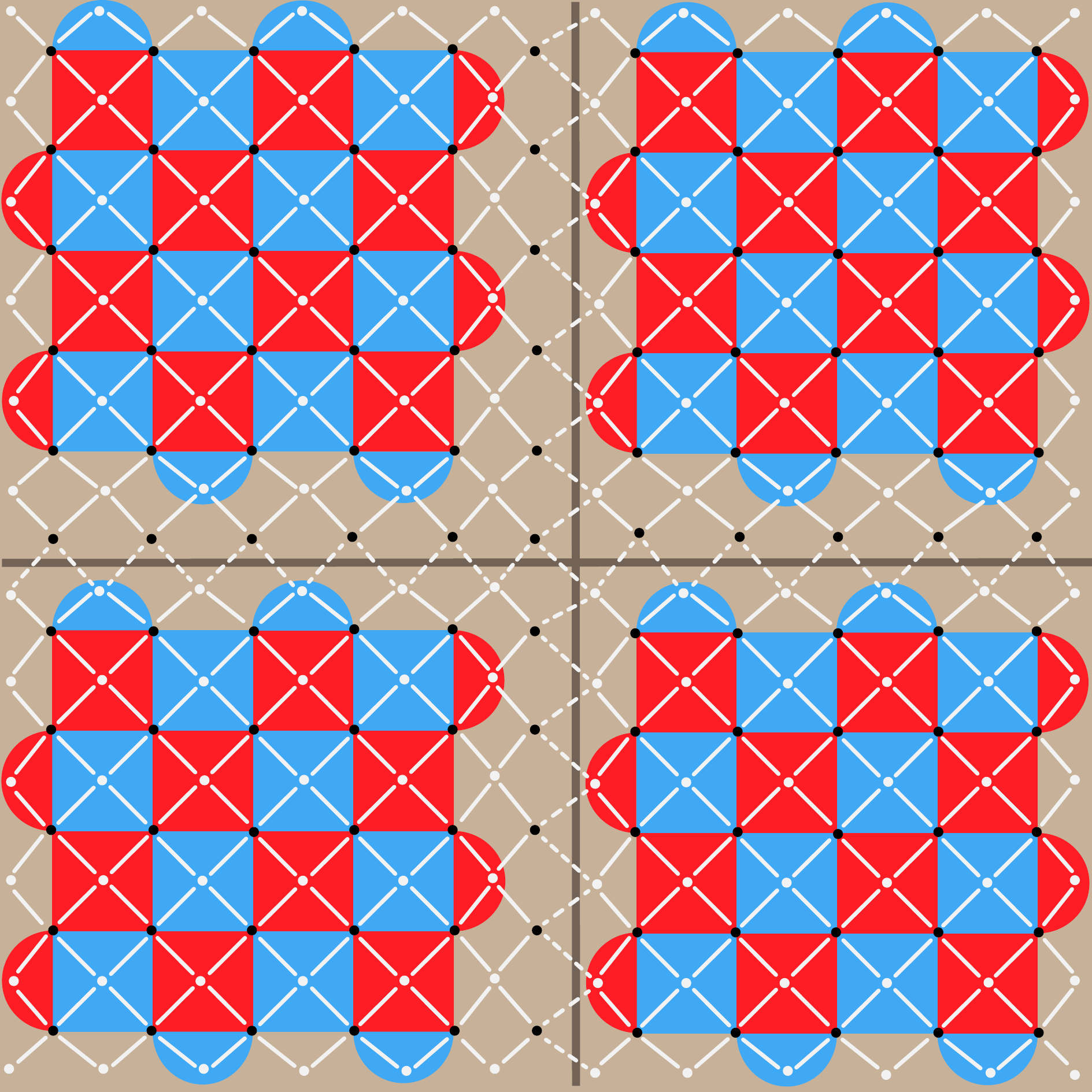}
  \caption{Schematic of a chiplet architecture.}
  \label{fig:chiplet}
\end{figure}

The modular architecture we simulate in this project is an array of chiplets similar to the ones in \cite{kate_chiplet}. We allocate one patch of surface code on each chiplet, as in Fig. \ref{fig:chiplet}. The qubits on adjacent chiplets can communicate via the inter-chip links (shown in dashed lines), but these links are currently $\sim 3X$ worse than on-chip links\cite{kate_chiplet}. Since no patch is defined across multiple chiplets, the inter-chip links are not used when a patch is idle. We assume the physical qubits on each chiplet has the grid connectivity, which is the one that naturally supports the surface code.

In Fig. \ref{fig:chiplet}, we show a design that allows one to swap the assignment of the data and syndrome qubits on a chiplet by rotating the chiplet by 180\textdegree. When a chiplet contains more faulty syndrome qubits than faulty data qubits, this will likely improve the quality of the code, because faulty syndrome qubits generally cause more significant drop in fidelity. In Sec. \ref{sec:yield_results}, we will evaluate how much advantage this degree of freedom translates to.

\subsection{Post-selection criterion: assessing the quality of defective chiplets}\label{sec:indicators}

When building a modular device, one has the opportunity to select the chips with better quality, and arrange them in a way that maximizes the fidelity. In \cite{kate_chiplet}, only the defect-free chiplets are kept, and then they are combined in a way that avoids frequency collisions along inter-chip links. When the goal is to support the surface code, we need different post-selection criteria. The chiplets that support higher-quality surface code patches should be kept, and they should be arranged to ensure that they can communicate at full code distance via lattice surgery.

For the purpose of selecting and arranging chiplets, we need to find good indicators for the ability of a defective chip to support a good surface code encoding and lattice surgery. This is because in a realistic setting, it might be impractical to experimentally measure the fidelity of surface code patches encoded on each chiplet before deciding which ones to use. Experimentally testing the fidelity of lattice surgery operations between patches on different chips is even less practical, since it requires repeatedly connecting and disconnecting chiplets to iterate through different combinations. When the target logical error rate (LER) is tiny, the cost of running simulations (e.g. a memory experiment) to estimate fidelity is also formidable. This is because when the LER is small, it takes too many shots to observe enough instances of logical errors.


To investigate the relevance of different figures of merit for many sample chiplets, we need a way of evaluating the quality of individual chiplets. We adopt Eqn.~(\ref{Eqn:FigureOfMerit}) to devise one such quantity. Specifically, we look to find the exponent $\alpha d$ of this expression.
To obtain this number, for each sample chiplet, we evaluate the logical failure rate as a function of $p$ for values of $ 5\times 10 ^{-4}\leq p \leq 2\times 10 ^{-3}$ where logical failure rates are determined using Monte Carlo methods. This is a typical regime where the defect-free surface code is studied~\cite{fowler2012surface}. The value $\alpha d$ is the gradient of the logical failure rate shown as a function of $p$ plotted with logarithmic axes. As such we will refer to this value as `the slope'. We show logical failure rates plotted as a function of $p$ in Fig.~\ref{fig:straight_lines_log}. The straight lines given in the plot indicate that we are sampling in a low $p$ regime where Eqn.~(\ref{Eqn:FigureOfMerit}) is valid.

\begin{figure}
  \centering
  \includegraphics[width=0.4\textwidth]{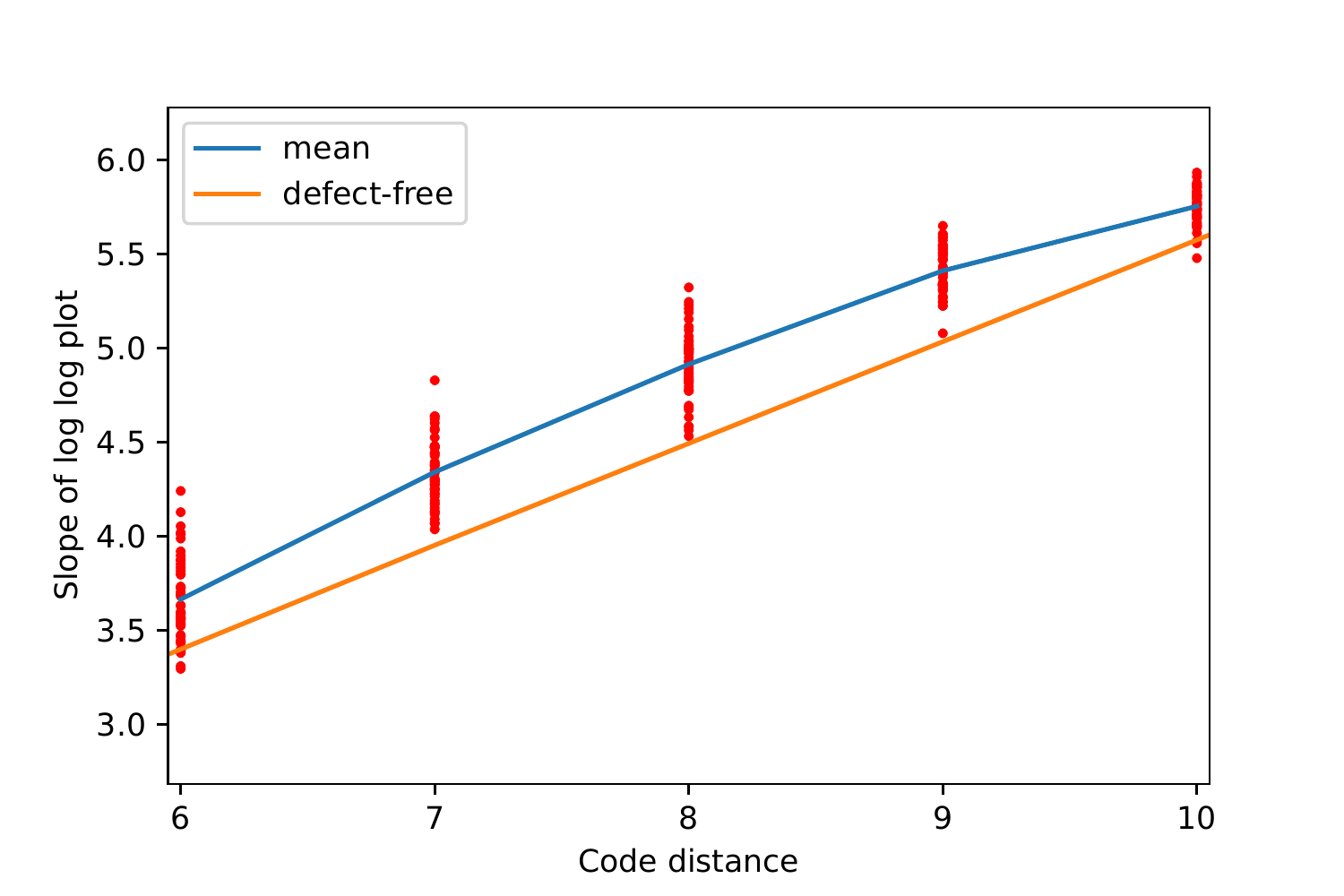}
  \caption{Slopes of the log-log LER v.s. $p$ plots, from randomly sampled defective rotated surface code patches with $l=11$. For each value of $d$, 50 defective patches are sampled, with the same probability for link failure and qubit failure.}
  \label{fig:slopes_scatterplot}
\end{figure}

We explored various possible indicators including $d$ of the defective patches, the total number of qubits that get disabled (a quantity that is generally higher than the number of faulty qubits), the size of the largest cluster of disabled qubits. The code distance of a defective surface code patch is the best indicator we find (Fig. \ref{fig:slopes_scatterplot}). Although \cite{dan2017} suggests that defective patches are outperformed by defect-free patches with the $d$, their data (Fig. 14 in \cite{dan2017}) only supports this claim for physical error rates $\geq 3\times 10^{-3}$. Instead, we find that the defective patches generally have higher slopes than the defect-free patches with the same $d$ (Fig. \ref{fig:slopes_scatterplot}). This means although the defective patches perform worse than the defect-free counterparts at higher $p$, they generally perform better at lower $p$.

\begin{figure}
  \centering
  \includegraphics[width=0.47\textwidth]{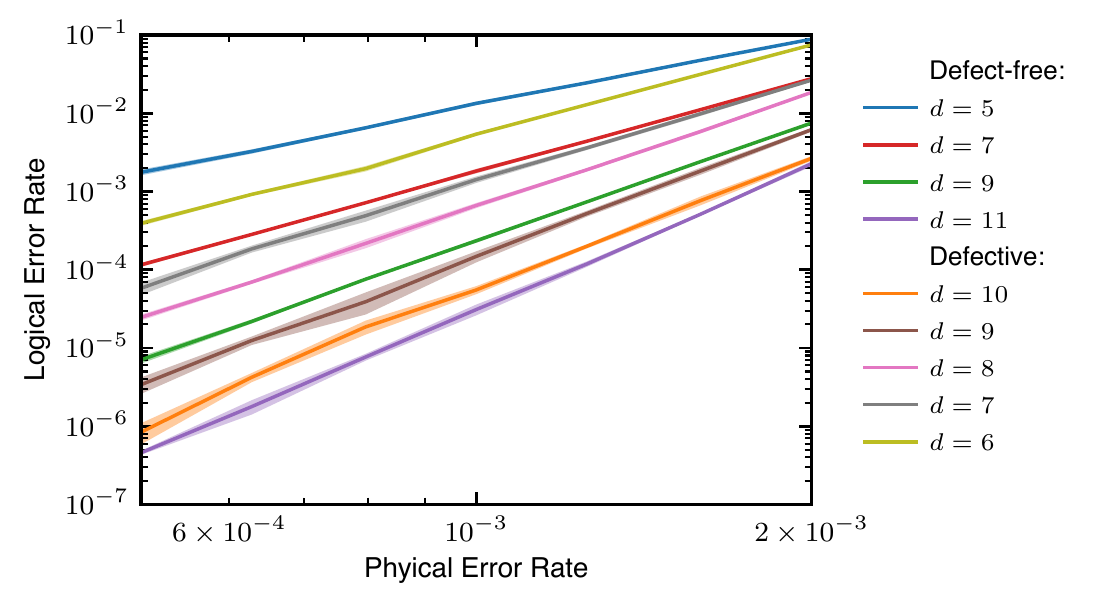}
  \caption{Logical error rate v.s. physical error rate at low physical error rates ($5\times 10^{-4}$ to $2\times 10^{-3})$, for defect-free patches of rotated surface code, and examples of defective patches with $l=11$. The shaded regions represent the 95\% confidence intervals for each value.}
  \label{fig:straight_lines_log}
\end{figure}

\begin{figure}
  \centering
  \includegraphics[width=0.46\textwidth]{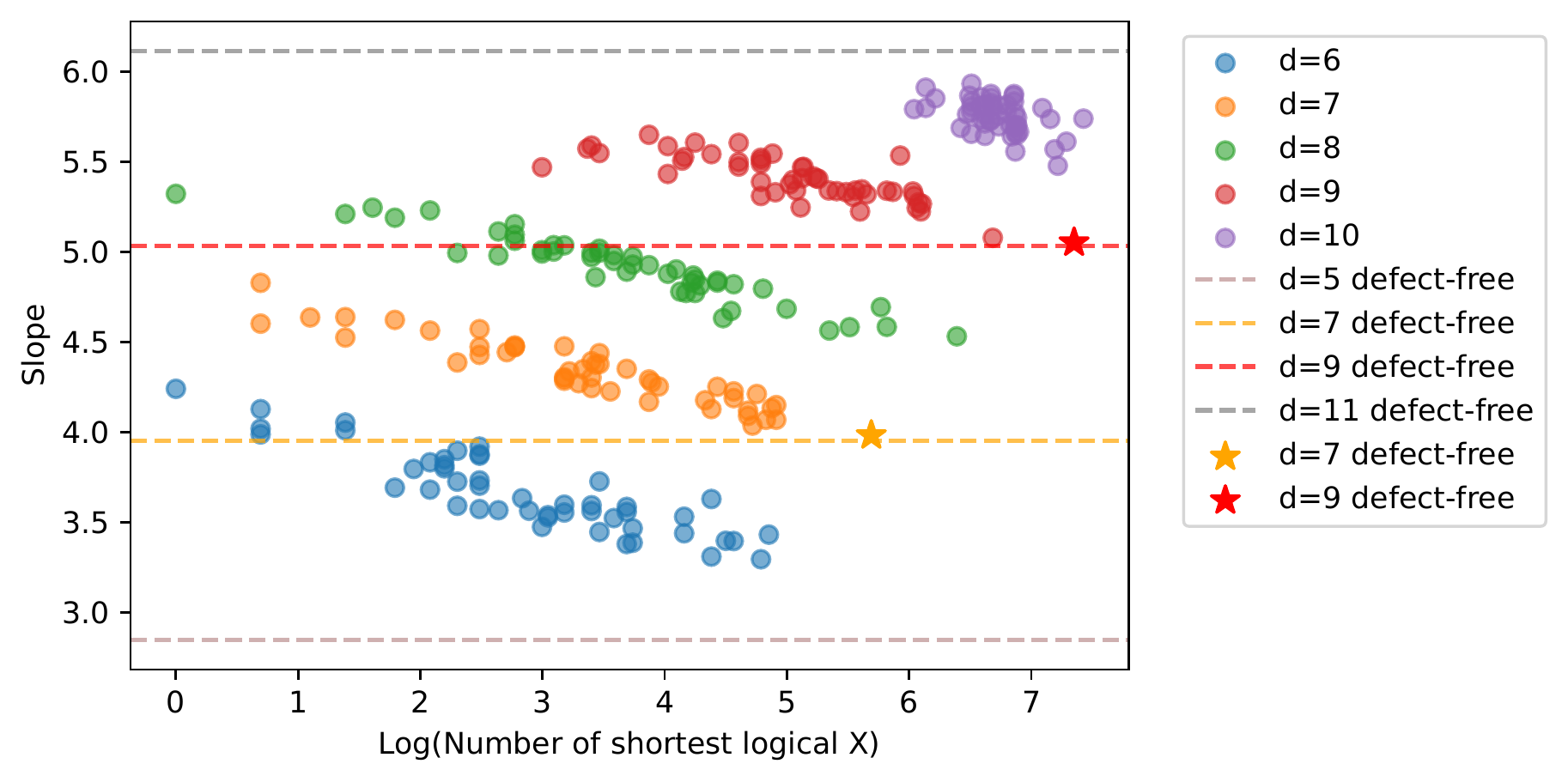}
  \caption{Slopes of LER v.s. $p$, from the same defective patches used in Fig. \ref{fig:slopes_scatterplot}, grouped by $d$ and plotted against the log of the number of shortest logical X.}
  \label{fig:num_shortest_paths}
\end{figure}

To explain the variation among patches with the same $d$, we identified a secondary indicator, the number of unique weight $d$ logical operators. In other words, it is the number of different ways that a logical failure can occur with $d$ physical errors. It can be evaluated efficiently with a modified version of breadth-first search on a graph where the nodes are the physical qubits on a surface code. As shown in Fig. \ref{fig:num_shortest_paths}, this helps to identify the outliers that significantly overperforms or under-performs compared to the defective patches with the same $d$. This indicator also helps us to understand why defective patches generally outperform defect-free counterparts with the same $d$: a defect-free patch has more symmetry in its shape so it has a large number of unique minimum-weight logical operators. 

\begin{figure}
  \centering
  \includegraphics[width=0.45\textwidth]{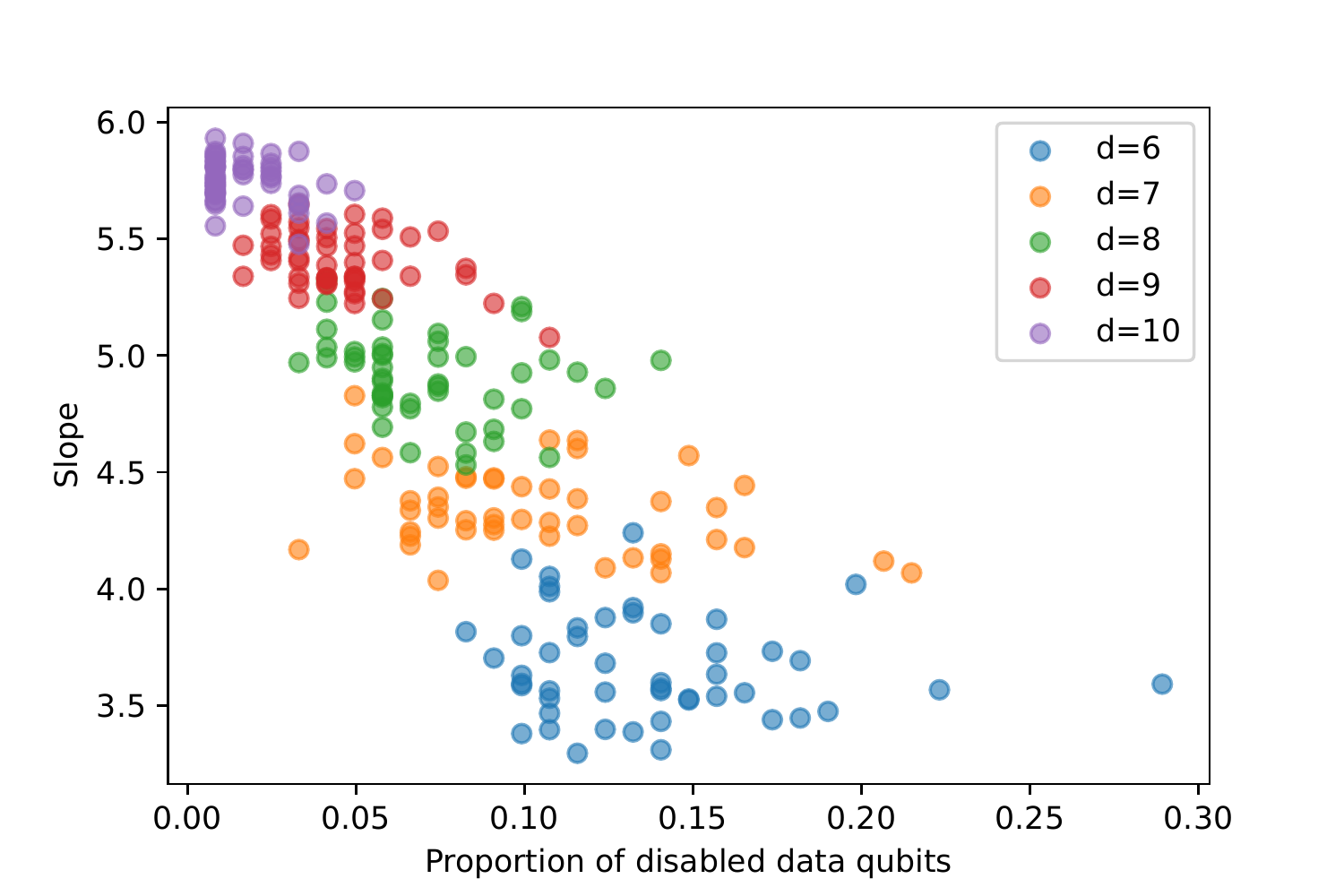}
  \caption{Slopes of LER v.s. $p$, from the same defective patches used in Fig. \ref{fig:slopes_scatterplot}, grouped by $d$ and plotted against the proportion of disabled physical qubits on a patch.}
  \label{fig:prop_disabled_qubs}
\end{figure}

\begin{figure}
  \centering
  \includegraphics[width=0.45\textwidth]{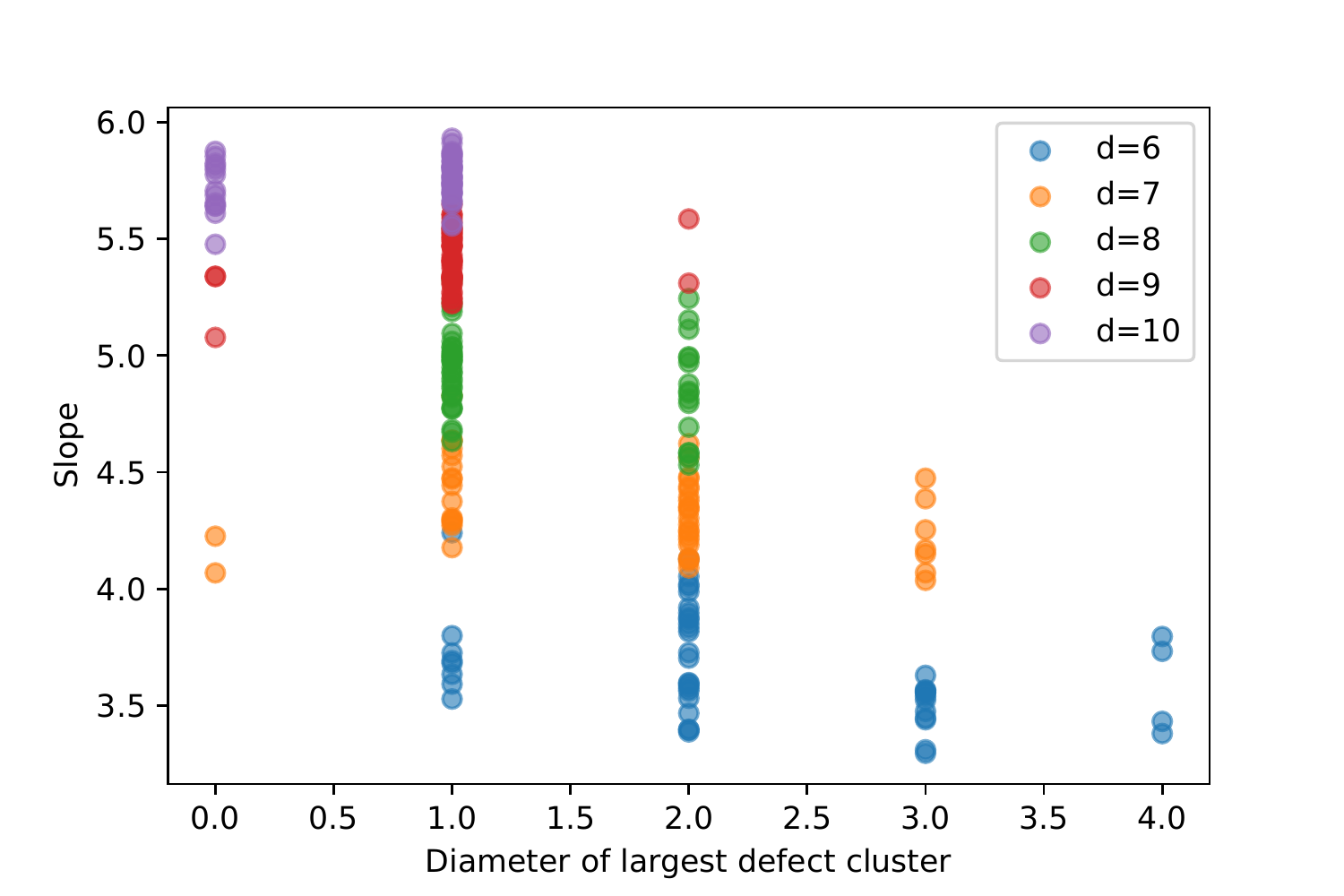}
  \caption{Slopes of LER v.s. $p$, from the same defective patches used in Fig. \ref{fig:slopes_scatterplot}, grouped by $d$ and plotted against the diameter of largest cluster of disabled qubits.}
  \label{fig:size_largest_cluster}
\end{figure}

In Fig. \ref{fig:prop_disabled_qubs} and \ref{fig:size_largest_cluster}, we evaluate two other indicators that we tested. The size of the largest defect cluster does not help predict the slope. The proportion of disabled data qubits is inversely correlated with the slope, but does not provide extra information that one cannot tell from the $d$.

\begin{figure}
  \centering
  \includegraphics[width=0.43\textwidth]{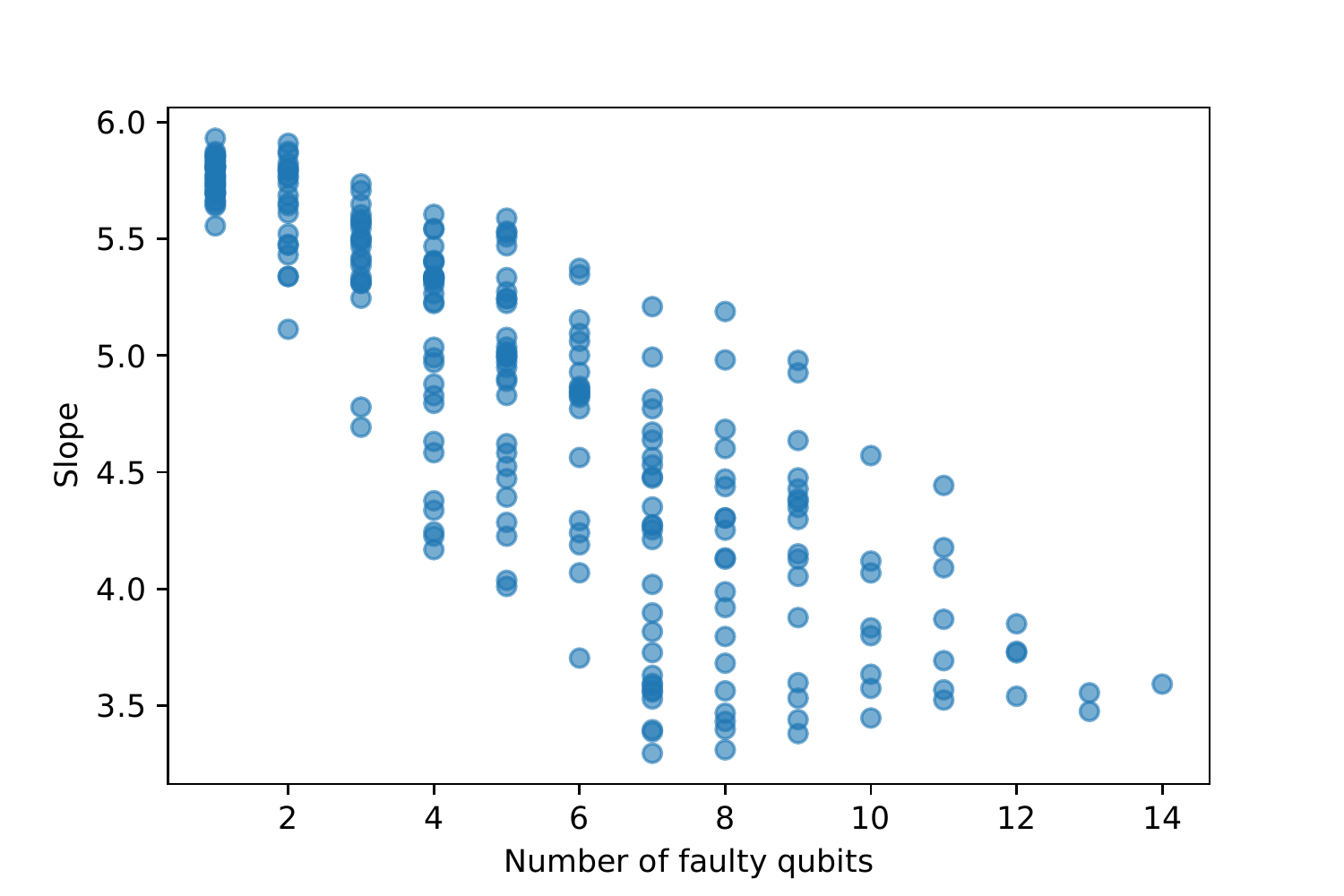}
  \caption{Slopes of LER v.s. $p$, from the same defective patches used in Fig. \ref{fig:slopes_scatterplot}, plotted against the number of faulty qubits on a patch.}
  \label{fig:baseline_indicator}
\end{figure}

\begin{figure}
  \centering
  \includegraphics[width=0.43\textwidth]{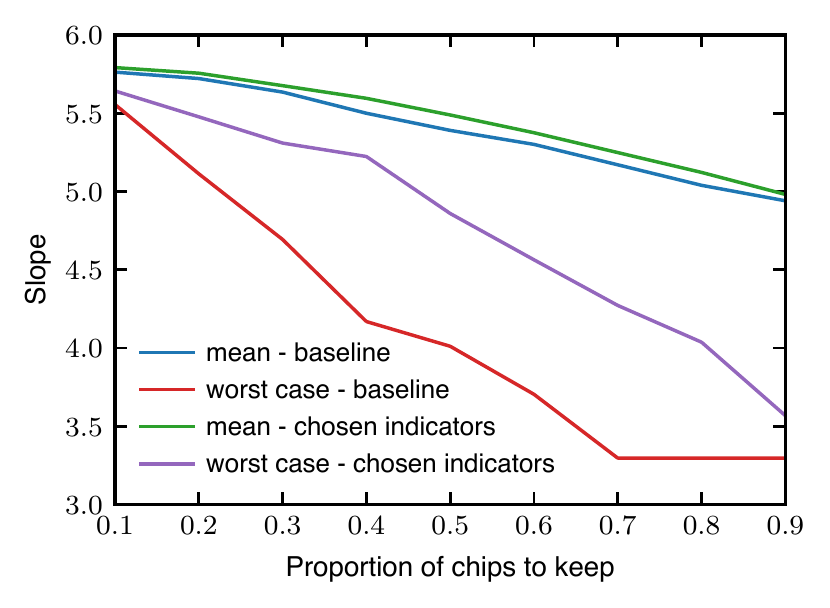}
  \caption{Mean and worst slopes of selected patches, when the proportion selected is varied. The baseline only uses the number of faulty qubits (Fig. \ref{fig:baseline_indicator}); the "chosen indicators'' use $d$ as primary indicator and the number of shortest logical operators to break ties.}
  \label{fig:compare_indicators}
\end{figure}

Now, we compare our post-selection criterion against a baseline indicator, the number of faulty qubits on a chiplet. Although there is a visible negative correlation between this quantity and the slope in Fig.~\ref{fig:baseline_indicator}, it is not as effective as the indicators we choose in Fig.~\ref{fig:compare_indicators}.

\section{Impact on resource overhead}
In this section, we show how design choices affect the resource overhead of building a large quantum computer to support a surface code.

\subsection{Resource overhead and sensitivity analysis}\label{sec:yield_results}
In this section, we show how fabrication errors increase the resource overhead of the surface code, and show that a modular architecture design and super-stabilizers successfully mitigate the cost.

\begin{figure}
\centering
\subfloat[]{%
  \includegraphics[clip,width=0.4\textwidth]{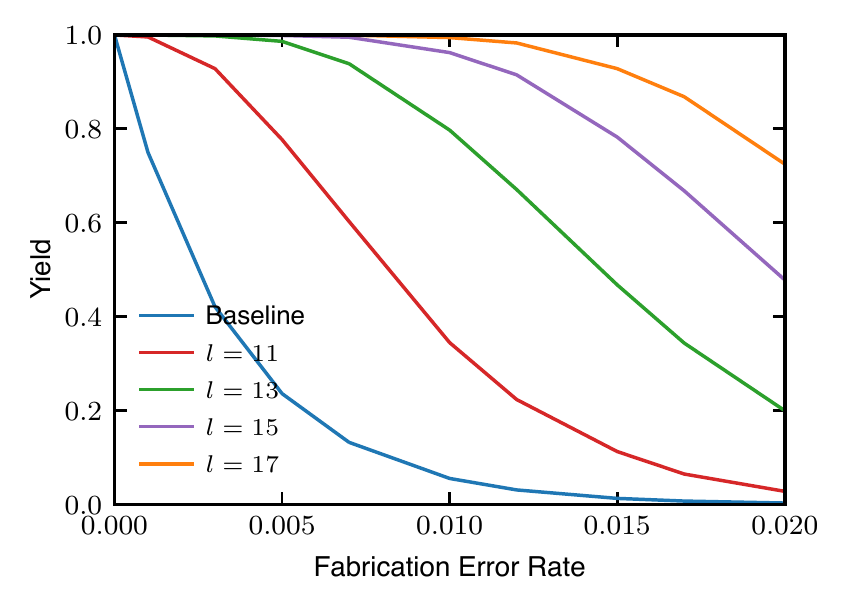}%
  \label{subfig:yield_fixed_target9_yield}
}
\qquad
\subfloat[]{%
  \includegraphics[clip,width=0.4\textwidth]{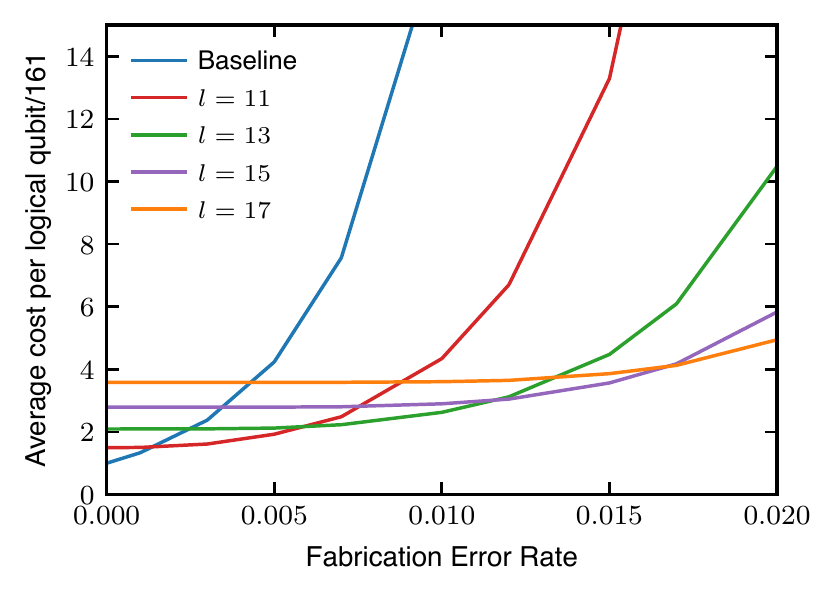}%
  \label{subfig:yield_fixed_target9_cost}
}
\caption{Defective links only. (a) Proportion of chiplets that support a rotated surface code patch that performs as well as a defect-free patch of distance 9, evaluated with the two metrics in \ref{fig:num_shortest_paths}. (b) As for (a) but showing the average number of fabricated physical qubits per logical qubit scaled by the number in the no-defect case.}
\label{fig:yield_fixed_target9}
\end{figure}

When the goal is to match the fidelity of the $d=9$ defect-free patch, we have the choice of using chiplets of width 9, 11, or larger. What chiplet size is more resource-efficient? If we make larger chiplets, each patch has a higher $l$. Then under the same fabrication error rate, we expect more chiplets to meet the standard. With smaller chiplets, we have a lower yield, and for the baseline where $l=9$, we cannot tolerate any defects. But each larger chiplet is made with more resources, which we quantify as the number of physical qubits. In Fig. \ref{fig:yield_fixed_target9}(a), we show the yields, and in (b), we show the average number of fabricated physical qubits for a logical qubit, which is obtained by dividing the number of qubits on each patch by the yield. The simulation is run with the model that only has faulty links, and each data point is collected from a 10000-shot simulation. From Fig. \ref{fig:yield_fixed_target9}(b) we can tell that below a fabrication error rate of $\sim0.1\%$, we should choose the baseline. From $\sim0.1\%$ to $\sim0.6\%$ and from $\sim0.6\%$ to $\sim1.1\%$, we should choose $l=$11 and 13 respectively. When the fabrication error rate is above $\sim1.1\%$, we should choose $l=15$ or higher. The overhead factor of the baseline approach rises out of the figure at higher defect rates. It is 18X and 336X respectively at 1\% and 2\% defect rates. 

\begin{figure}
\centering
\subfloat[]{%
  \includegraphics[clip,width=0.4\textwidth]{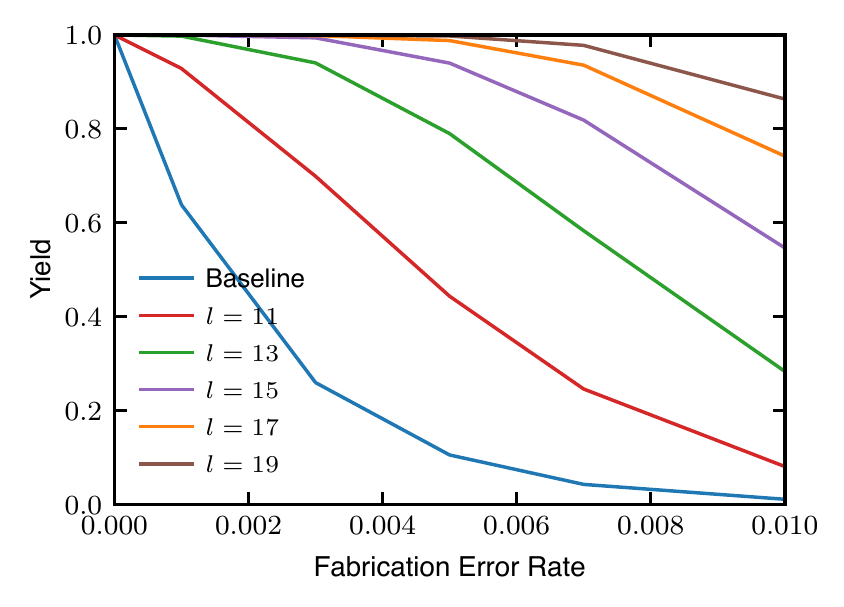}%
  \label{subfig:yield_tunable_target9_yield}
}
\qquad
\subfloat[]{%
  \includegraphics[clip,width=0.4\textwidth]{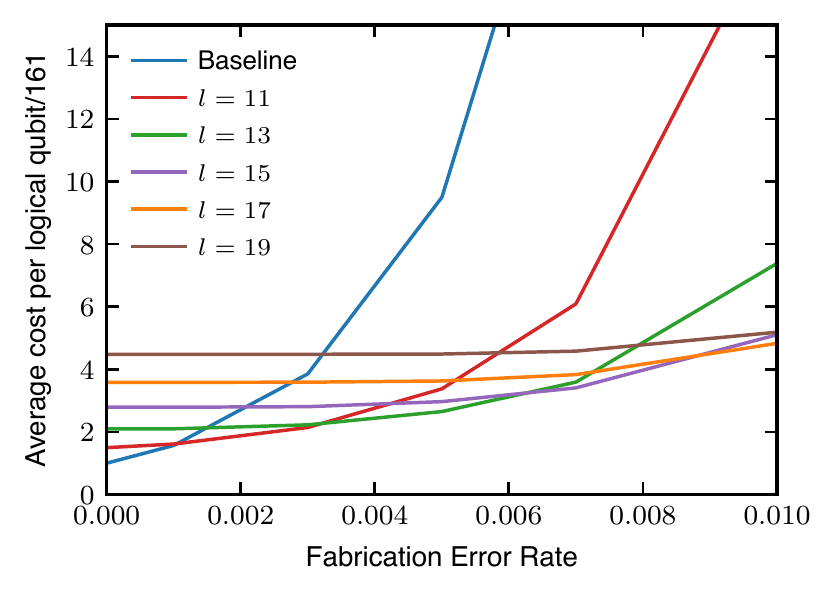}%
  \label{subfig:yield_tunable_target9_cost}
}
\caption{Links and qubits are assigned faulty at the same rate. (a) Proportion of chiplets that support a rotated surface code patch that performs as well as a defect-free patch of distance 9, evaluated with the two metrics in \ref{fig:num_shortest_paths}. (b) The same as for (a) but showing the average number of fabricated physical qubits per logical qubit scaled by the number in the no-defect case.}
\label{fig:yield_tunable_target9}
\end{figure}

When we use the model where links and qubits have the same defect rate (Fig. \ref{fig:yield_tunable_target9}), instead of the link-defect-only model, the yields are lower than in Fig. \ref{fig:yield_fixed_target9} and the advantage of using larger $l$ start from a lower defect rate. At a 1\% defect rate, the overhead factor of the baseline approach ($l=9$) is 91X.

For most of the paper, we focus on the fidelity of an individual patch. Here, we briefly discuss how certain deformations on the boundary would result in a drop in code distance during lattice surgery, and evaluate the cost of avoiding such chiplets.

\begin{figure}
  \centering
  \includegraphics[width=0.4\textwidth]{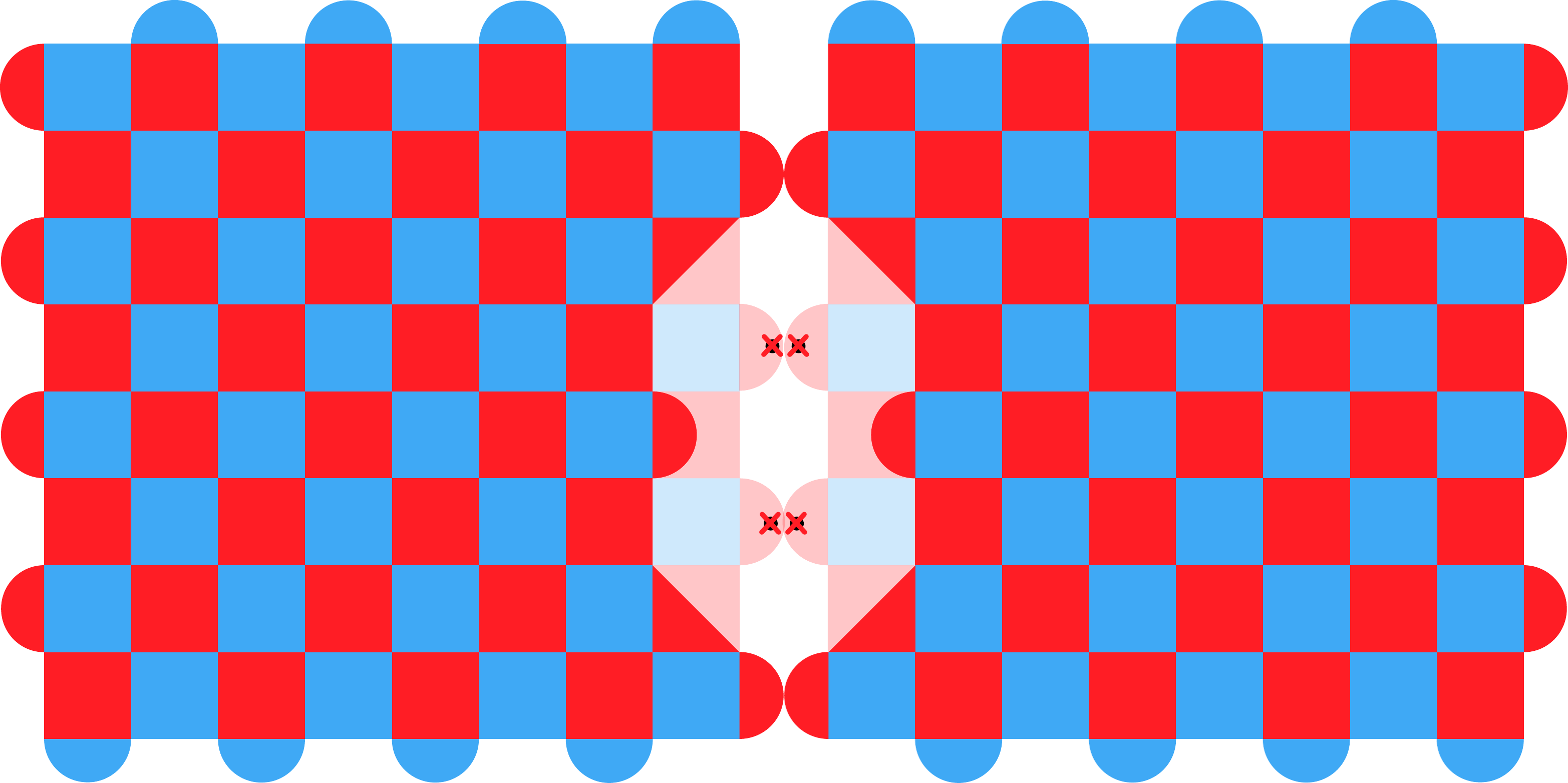}
  \caption{An example where the code distance drops after a merge.}
  \label{fig:ls_distance_drop}
\end{figure}

Lattice surgery involves merges and splits between patches of the planar surface code. In Fig. \ref{fig:ls_distance_drop} we show an example case where a deformed boundary only leads to a small decrease in the distance of the individual patch, but causes a larger decrease of code distance after a merge. In this example, the two merging edges are deformed at the same place. When the deformations are not aligned, there can be a greater drop in code distance after the merge. A low-distance merge has lower fidelity, so when this type of patch is used, the compiler should try to schedule lattice surgery operations on its other edges. Then, the programs would be compiled to more layers. Alternatively, one can avoid using patches with such an edge, which may result in a lower yield.

Note that we did not run simulation to compute the fidelity of lattice surgery operations between defective patches. Therefore, it is only our speculation that the code distance of the merged patch is sufficient to predict the fidelity of the logical operations.

\begin{figure}
  \centering
  \includegraphics[width=0.4\textwidth]{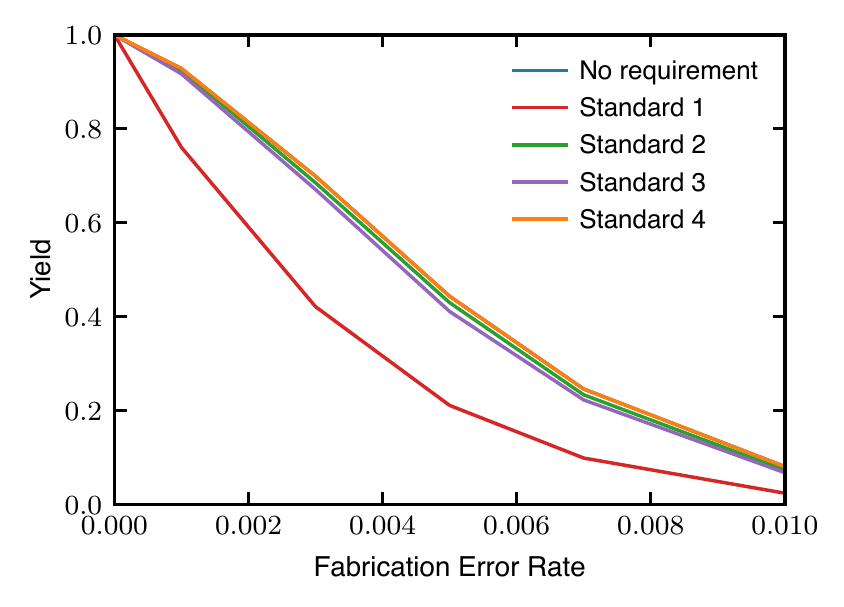}
  \caption{The change in yield after imposing different standards on boundaries of patches. Standard 1: No deformation on any boundary; standard 2: at least 2 boundaries of different types have no deformation; standard 3: all 4 boundaries support lattice surgery without decreasing distance; standard 4: at least 2 boundaries of different types support lattice surgery without decreasing distance.}
  \label{fig:boundary_condition}
\end{figure}

In Fig. \ref{fig:boundary_condition}, we show how the yield changes after a boundary constraint is imposed. We have two types of boundary constraints for an edge of a surface code patch: (a) where we are free of deformations (in this case we do not need to form any new super-stabilizer during lattice surgery), and (b) where the total width of deformations along the edge is not enough to decrease the code distance after a merge. Then, we have the choice of imposing the constraint on (c) all four edges of a patch, or (d) on only two edges (at least one X-edge and at least one Z-edge, for the convenience of scheduling lattice surgery operations). From these, we get four different boundary constraints. The yield drops significantly only when we impose the strictest constraint (standard 1, or a and c). The drop is negligible for standard 4, when we impose (b) and (d). When we impose standard 2 or 3, the drop is visible but small. Given the results, we should apply standard 3 if we are willing to form new super-stabilizers along the merging/splitting edges; if not, we should apply standard 2.

\begin{figure}
  \centering
  \includegraphics[width=0.4\textwidth]{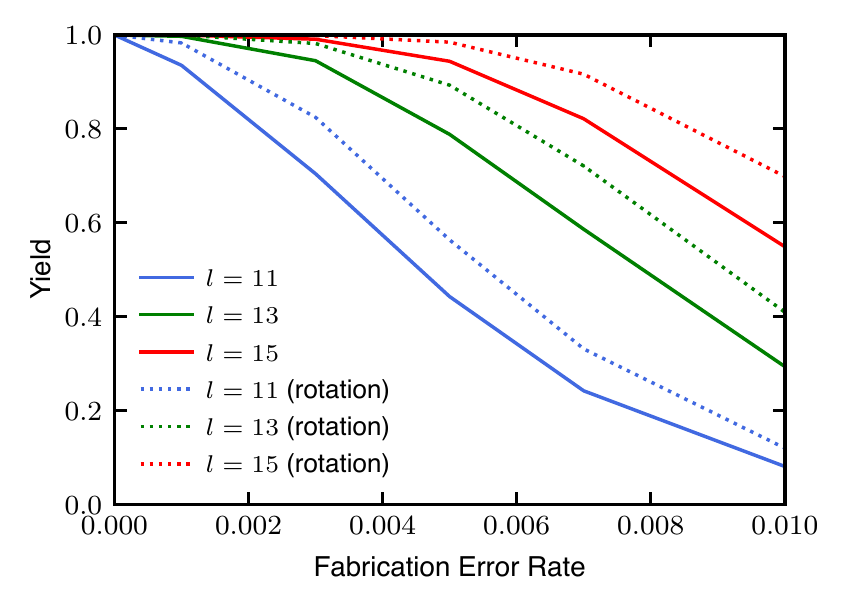}
  \caption{Improvement in yield when there is freedom to rotate the chiplets. Links and qubits are assigned faulty at the same rate.}
  \label{fig:benefit_rotation}
\end{figure}

The way we propose to allocate a surface code patch on a chiplet (in Sec \ref{sec:describe_arch}) allows the freedom to swap the assignment of data/syndrome qubits by a 180\textdegree rotation. Alternatively, this freedom can be achieved by translating the position of the logical qubit by one physical qubit, but this translation needs to be coordinated so that the adjustments to neighboring patches do not conflict. We observe an improvement in yield when we have such a freedom (Fig. \ref{fig:benefit_rotation}), when qubit defects are present.
Some techniques to reduce leakage errors involve swapping data and ancilla qubits~\cite{mcewen2021removing}, which might not work well with this design. Handling leakage errors is outside the scope of this paper, however.

\begin{figure}
\centering
\subfloat[]{%
  \includegraphics[clip,width=0.4\textwidth]{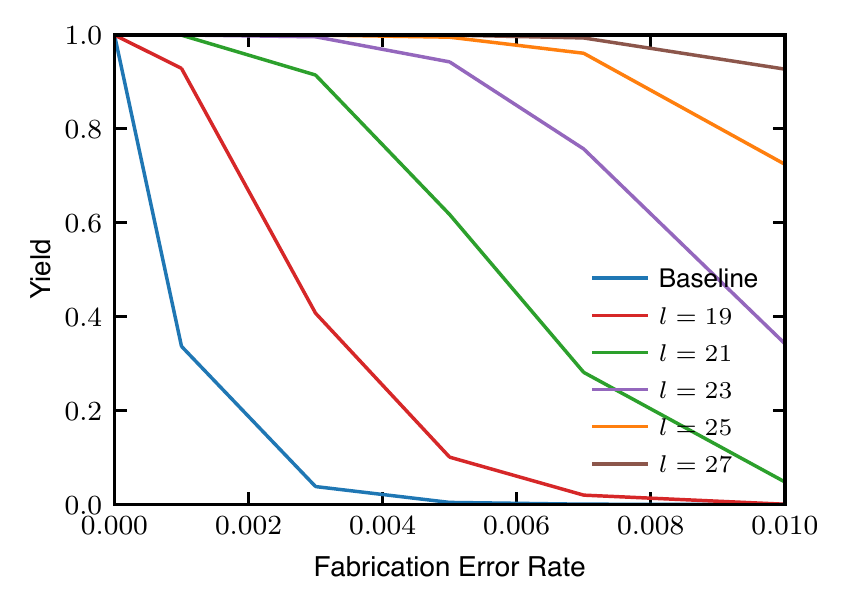}%
  \label{subfig:yield_fixed_target17_yield}
}
\qquad
\subfloat[]{%
  \includegraphics[clip,width=0.4\textwidth]{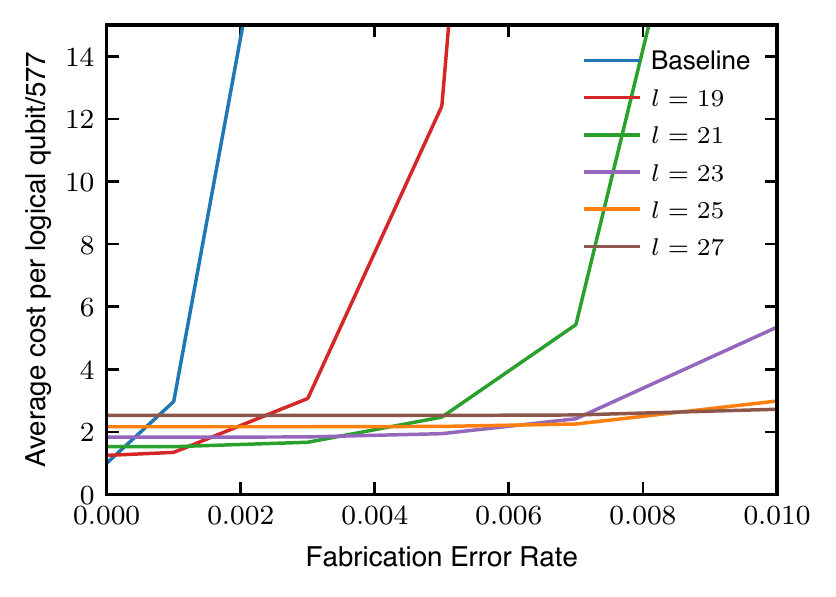}%
  \label{subfig:yield_fixed_target17_cost}
}
\caption{Yields for larger chiplets. Defective links only. (a) Proportion of chiplets that support a rotated surface code patch that performs as well as a defect-free patch of distance 17, evaluated with the two metrics in \ref{fig:num_shortest_paths}. (b) The same as for (a) but showing the average number of fabricated physical qubits per logical qubit scaled by the number in the no-defect case.}
\label{fig:yield_fixed_target17}
\end{figure}

In Fig. \ref{fig:yield_fixed_target17}, we show the cost of making higher-quality logical qubits. For this set of simulations, instead of matching the fidelity of the $d=9$ defect-free rotated surface code, the goal is to match the $d=17$ defect-free code. The trends we observe are qualitatively the same. Note that at 1\% defect rate, the factor of resource overhead from the $l=17$ defect-intolerant baseline is over 56000X. In fact, when the defect rate is fixed, the resource overhead increases exponentially with the number of physical qubits in a logical qubit.

\begin{figure}
\centering
\subfloat[Defective links only.]{%
  \includegraphics[clip,width=0.4\textwidth]{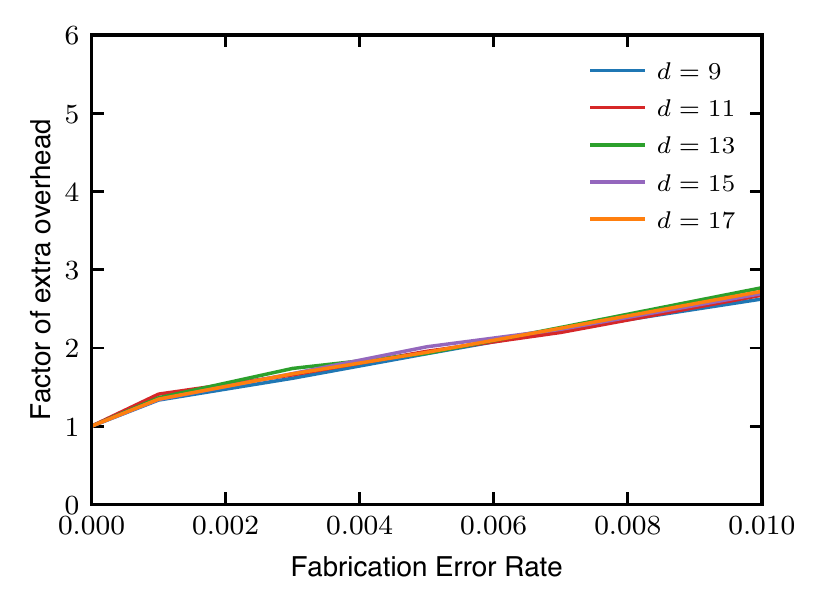}%
  \label{subfig:fixed_lowest_overhead}
}
\qquad
\subfloat[Defective links and qubits, without swapping data and syndrome qubits.]{%
  \includegraphics[clip,width=0.4\textwidth]{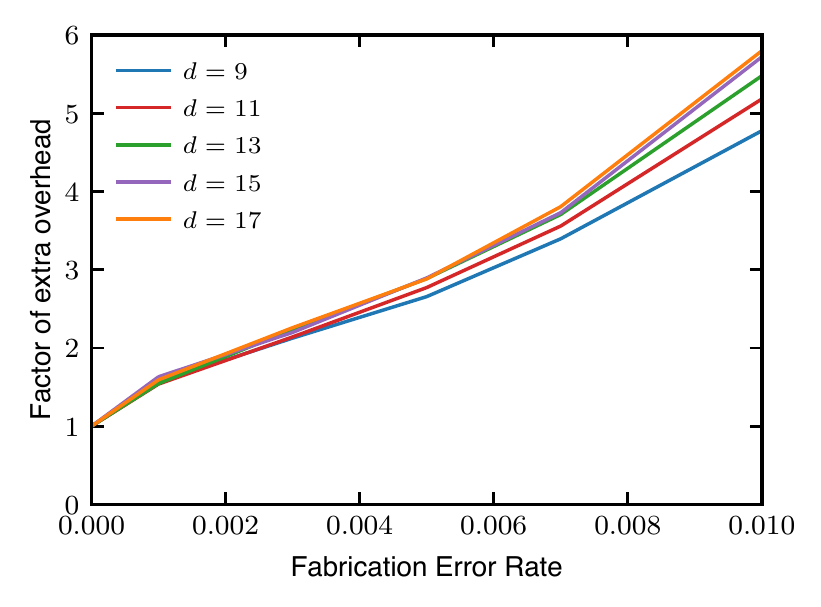}%
  \label{subfig:tunable_lowest_overhead}
}
\qquad
\subfloat[Defective links and qubits, with the option to swap the assignment of data and syndrome qubits.]{%
  \includegraphics[clip,width=0.4\textwidth]{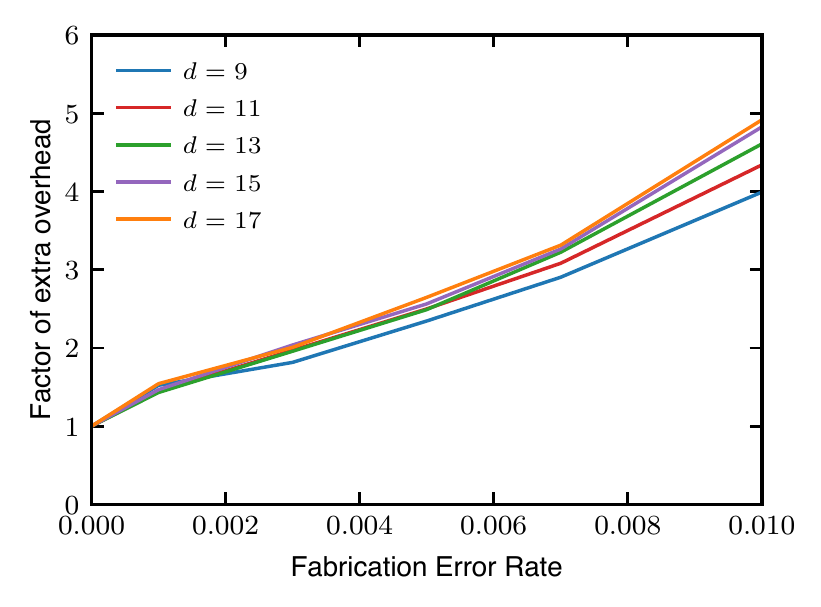}%
  \label{subfig:rot_lowest_overhead}
}
\caption{The extra resource overhead due to defects, for different target logical error rates. The y-axis is the average number of fabricated qubits for a logical qubit, scaled by the number in the ideal no-defect case. The target logical error rate is the fidelity of the defect-free rotated surface code of code distance $d$.}
\label{fig:lowest_overhead}
\end{figure}

On Fig. \ref{subfig:yield_fixed_target9_cost},\ref{subfig:yield_tunable_target9_cost}, and \ref{subfig:yield_fixed_target17_cost}, if we take the minimum of all curves at each fabrication error rate, we obtain the minimum extra resource overhead (due to defects) that can be achieved by the chiplet architecture considered in this work. In Fig. \ref{fig:lowest_overhead}, we show how this factor is affected by the fabrication error rate and the target code fidelity. When the fabrication error model consists of defective links only, the curves for different target fidelity coincide. It is $\sim 2X$ at a $0.5\%$ defect rate, and below $3X$ at $1\%$ defect rate. When we model both defective qubits and links, the curves coincide at low defect rate and diverge a small amount at higher defect rate. The factor of overhead is $\sim 3X$ at a $0.5\%$ defect rate and 5X to 6X at $1\%$ for the range of fidelity targets we set. With the freedom to swap the assignment of data and ancilla qubits, the resource overhead can be lower, as shown in Fig. \ref{subfig:rot_lowest_overhead}.

Fig. \ref{fig:lowest_overhead} shows that, even with a method to implement QEC in the presence of fabrication errors, it is still important to reduce the defect rate by adopting improved designs for qubits or fabrication. Limiting the factor of overhead to below 2X requires a defect rate below $\sim 0.5\%$ for the link-defect only model, and below $\sim 0.3\%$ for the model with both defective links and qubits. 

\subsection{Limit of the monolithic architecture}\label{sec:monolithic}
The results in Sec. \ref{sec:yield_results} demonstrate that the ability to implement the surface code on defective grids is necessary for containing its resource overhead, by comparing against a baseline design that uses modular chiplets but only accepts defect-free surface code patches. What if we choose a monolithic architecture but accept defective patches?

The monolithic design does not allow for post-selection of chiplets, thus the resulting chip will contain regions that cannot support logical qubits that meet the requirement. Furthermore, without the freedom to arrange the chiplets, one cannot avoid the situation in Fig. \ref{fig:ls_distance_drop} and ensure that each patch can interact with neighboring patches with high fidelity.

As the next section (Section \ref{sec:shor}) will demonstrate through a case study, if the suboptimal regions of a monolithic device are used, then they will lower the application fidelity. Alternatively, one can restrict the compiler to only use the patches that meet the requirement. However, this constraint effectively reduces the connectivity between logical qubits. This will increase the number of time steps that it takes to run the program, hence reduce the program fidelity.
\subsection{Resource overhead and fidelity estimation for an example application}\label{sec:shor}

In Section \ref{sec:yield_results}, the quantity we use for evaluating each approach is the factor of extra resource overhead relative to the ideal defect-free case. In this section, we estimate the resource cost and application fidelity of an example application in the presence of defects.

The application we choose is Shor's algorithm applied to 2048-bit integers, whose implementation with surface code (in the no-defect setup) is optimized and analyzed in ~\cite{gidney2021factor}. It requires a $226\cdot 63$ grid of distance-27 surface code patches, and about 25 billion surface code cycles, according to ~\cite{gidney2021factor}.

In Table \ref{table:lowp} and \ref{table:highp},  we show cost estimates for building a modular device that supports the application, at a defect rate of $0.1\%$ and $0.3\%$ respectively (on both qubits and links). To compute the cost of the super-stabilizer approach, we use the steps in Section \ref{sec:yield_results}, and find the optimal choice of $l$ that minimizes the factor of resource overhead for a target code distance of 27. The no-defect baseline is for the ideal case where no defects arise, and the defect-intolerant baseline is for the design choice to only use defect-free chiplets. The results show that the defect-intolerant approach makes the already tremendous resource requirement for the algorithm orders of magnitude higher, while the super-stabilizers help lower the factor of resource overhead to a small number (45X better for the $0.1\%$ defect rate and more than $10^5$X better when the defect rate is $0.3\%$). This example also demonstrates the importance of reducing the defect rate. When the defect rate is increased from $0.1\%$ to $0.3\%$, the cost increases by $40\%$ even when the super-stabilizers are applied.
\begin{table}[ht!]
\centering
\begin{tabular}{c c| c c}
 \hline
  & No-defect & \thead{Defect-\\ intolerant} & \thead{Super-\\stabilizer} \\ [0.5ex] 
 \hline
 $l$ & 27 & 27 & 33 \\ 
 Yield & 100\% & 1.4\% & 94.5\% \\ 
Overhead & 1 & 71.32 & 1.58 \\
 Qubits & $2.1\times 10^7$ &$1.5\times 10^9$  & $3.3\times 10^7$ \\ [1ex] 
 \hline
\end{tabular}
\caption{Resource estimation for building a device that supports a $226\cdot 63$ grid of distance-27 surface code patches, for a defect rate of $0.001$ on both qubits and links. \textit{No-defect} is the ideal setting without fabrication defects, not an approach to handle defects. Both the \textit{defect-intolerant} and the \textit{super-stabilizer} approaches here post-select chiplets to make a modular device. \textit{Overhead} is the factor of resource overhead, determined by the yield and the size of each chiplet; \textit{Qubits} is the total number of physical qubits fabricated for the application. \label{table:lowp}}
\begin{tabular}{c c| c c} 
 \hline
 & No-defect & \thead{Defect-\\ intolerant} & \thead{Super-\\stabilizer} \\ [0.5ex] 
 \hline
 $l$ & 27 & 27 & 39 \\
 Yield & 100\% & $2.7\times 10^{-6}$ & 94.6\% \\ 
 Overhead & 1 & $3.67\times 10^5$ & 2.21 \\
 Qubits & $2.1\times 10^7$ & $7.6\times 10^{12}$ & $4.6\times 10^7$ \\ [1ex] 
 \hline
\end{tabular}
\caption{Same as Table \ref{table:lowp} but for a defect rate of $0.003$.\label{table:highp}}
\end{table}

\begin{figure}
\centering
\subfloat[$l=33$, defect rate at $0.1\%$ for both links and qubits.]{%
  \includegraphics[clip,width=0.4\textwidth]{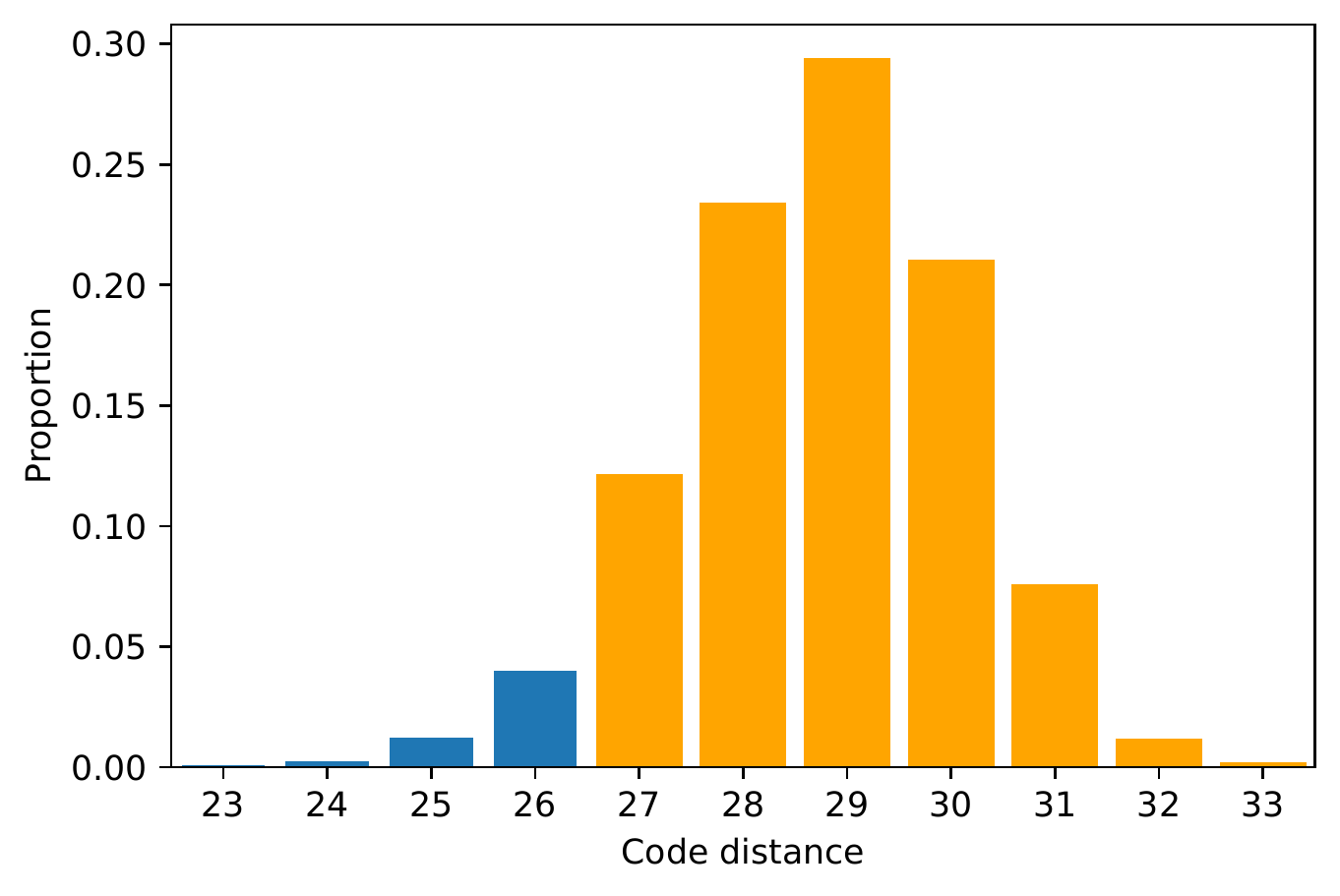}%
  \label{subfig:distance_distribution_l33}
}
\qquad
\subfloat[$l=39$, defect rate at $0.3\%$ for both links and qubits.]{%
  \includegraphics[clip,width=0.4\textwidth]{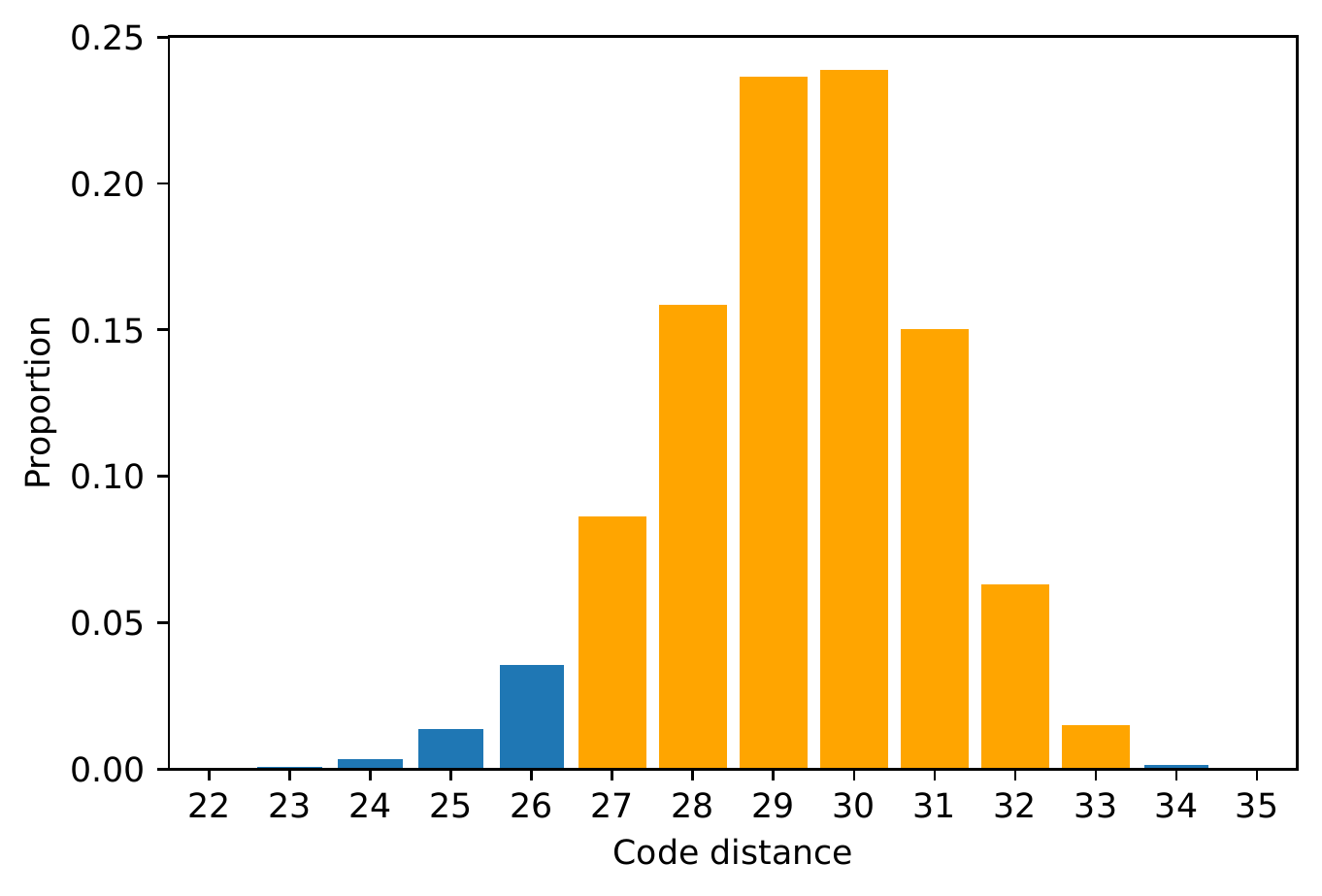}%
  \label{subfig:distance_distribution_l39}
}
\caption{Distribution of code distance. In orange: proportions of patches with $d \geq 27$; in blue: proportions of patches with $d < 27$. The results are obtained without reassigning data and syndrome qubits.}
\label{fig:distance_distribution}
\end{figure}

The fidelity of a large-scale fault-tolerant application can be roughly estimated with the topological error rate, as in Section 2.13 of ~\cite{gidney2021factor}. We follow their method to estimate the fidelity of the application, assuming the physical gate error on the device is $10^{-3}$. In the calculation, we account for the code distance distributions for the adapted surface code patches (see Fig. \ref{fig:distance_distribution}). The distributions are each obtained from a sample size of 10000. Recall that based on the results in Section \ref{sec:indicators}, code distance is the most important indicator for the performance of an adapted surface code patch. Furthermore, the logical error rate of the adapted surface code is generally lower than that of the defect-free patch with the same code distance. Therefore, using code distance to estimate the performance of each patch does not underestimate the failure rate for the super-stabilizer approach.

Estimates for the application fidelity are shown in Table \ref{table:fidelity_lowp} and \ref{table:fidelity_highp}. Note that the device from the ideal no-defect case (all patches are exactly distance 27) would have a fidelity of $\sim 73\%$. In the case where we use a modular device with super-stabilizers, all the patches are at least distance 27 and most of them have larger code distances. Therefore, the estimated application fidelity is higher, albeit at a higher resource cost than the ideal no-defect case. 

\begin{table}[ht]
\centering
\begin{tabular}{c c c c}
 \hline & baseline1 & baseline2 & \thead{Modular $\&$\\ super-stabilizer} \\ [0.5ex]
 \hline
 $l$ & $15\sim17$ & $33\sim35$ & 33 \\ 
Overhead & 1.58 & 1.58 & 1.58 \\
 \thead{Estimated\\fidelity} & 0 & $79.9\%$ & $88.5\%$ \\ [1ex] 
 \hline
\end{tabular}
\caption{Application fidelity estimated with the topological error. Baseline1: modular, defect-intolerant. Baseline2: monolithic, uses super-stabilizers to handle defects. \label{table:fidelity_lowp}}

\begin{tabular}{c c c c}
 \hline & baseline1 & baseline2 & \thead{Modular $\&$\\ super-stabilizer} \\ [0.5ex]
 \hline
 $l$ & $11\sim13$ & $39\sim41$ & 39 \\ 
Overhead & 2.21 & 2.21 & 2.21 \\
 \thead{Estimated\\fidelity} & 0 & $76.1\%$ & $91.7\%$ \\ [1ex] 
 \hline
\end{tabular}
\caption{Same as Table \ref{table:fidelity_lowp} but for a defect rate of $0.003$.\label{table:fidelity_highp}}
\end{table}

We compare our approach against two baselines while holding the resource overhead constant. The first baseline's goal is to build a modular, defect-free device from defective qubits. For this baseline, we need to lower the resource overhead to be the same as the super-stabilizer approach. We do this by reducing the size of the chiplets. The factor of resource overhead for building this modular defect-free device from defective qubits (relative to the distance-27, ideal no-defect case) is 1.12 for $d=15$ and 2.09 for $d=17$, at a defect rate of $0.1\%$. To match the resource overhead of our approach (1.58), one would use a mix of $d=15$ and $d=17$ defect-free patches. However, these code distances are insufficient for the application; they both result in an estimated fidelity of effectively 0. Furthermore, when the defect rate is $0.3\%$, the defect-intolerant baseline can only afford patches of distance 11 and 13, which is even farther below the requirement.

The second baseline is the monolithic device with the super-stabilizers applied. There is no post-selection of chiplets for the monolithic device, so its resource overhead is lower if the same $l$ is used. In order to match the resource overhead factor of our approach, we increase $l$ for a proportion of the patches. For the case with a $0.1\%$ defect rate, we keep $53\%$ of the patches at $l=33$ and use $l=35$ for the rest. For the defect rate$=0.3\%$ case, we keep $47\%$ of the patches at $l=39$ and expand the rest to $l=41$. Our calculation shows that even with the increased code distance, the expected failure rate of this baseline is $\sim 1.8$X and $\sim2.9$X higher than the modular, super-stabilizer approach for the two cases. This is because the logical qubit patches with $d < 27$ contribute to the error rate on the monolithic device, negatively impacting fidelity, but they are discarded in the modular case. 
\section{What counts as a fabrication error}\label{sec:stability}

So far we have been using a simple model for defects, where each qubit or link is labeled directly as ``faulty'' or ``non-faulty''. In real life, there are scenarios where it is not clear whether a qubit should be viewed as faulty. For example, on a device with a average 2-qubit gate fidelity of $99.9\%$, should we disable a qubit that only supports 2-qubit gates with $97\%$ fidelity? If the cutoff is set too high, we lose too many physical qubits and suffer a decrease in code performance; if the cutoff is too low, the inferior qubits will also damage the code.

To identify cutoff fidelity values for labelling a qubit faulty, we need to compare the logical performance when we keep the faulty qubits against the results from disabling them.

We use the stability experiment \cite{Gidney2022stability} instead of the more standard memory experiment that we used in the previous sections. While the memory experiment quantifies how well a logical observable is maintained by the code, the stability experiment evaluates how well a logical observable can be moved (a capacity that is needed for logical operations). As explained in \cite{Gidney2022stability}, measurement errors can't cause logical errors in memory experiments except by creating confusion that hides the key data errors. Since errors caused by a faulty qubit look like repeated measurement errors, the memory experiment is unable to show the damage of faulty qubits.

\begin{figure}
  \centering
  \includegraphics[width=0.45\textwidth]{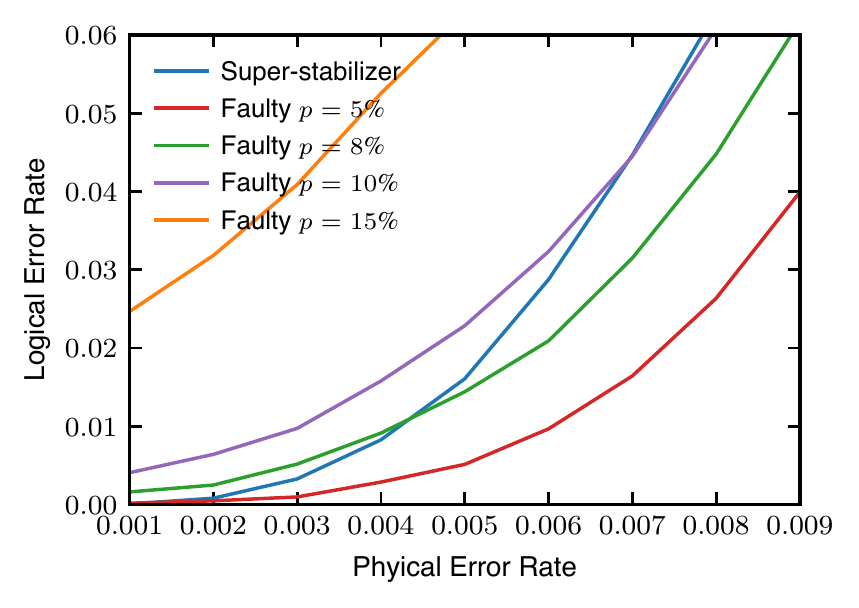}
  \caption{Stability experiment results from keeping/disabling a bad data qubit on a $d=5$ surface code. The x-axis is the physical error rate of the good qubits.}
  \label{fig:stability}
\end{figure}

In Fig. \ref{fig:stability}, we show the results from a stability experiment where the data qubit in the center of a $d=5$ surface code has higher error rate than the rest. We set the value of $p$, the two-qubit error rate of the worse qubit, to values from $5\%$ to $15\%$ (the other errors on it scale accordingly). Then, we compare the results against the one from disabling the worse qubit and using super-stabilizers around it. The figure shows that when $p$ of the bad qubit is above $\sim 10\%$, we should disable it regardless of the quality of the other qubits. When $p$ is below $5\%$, it is preferable to keep it unless the error rate on the other qubits is below the range in the plot. Finally, when $p=8\%$, we should disable the qubit if the error rate on the other qubits is below $\sim 0.45\%$. 

\section{Related work}

To the best of our knowledge,~\cite{kate_chiplet} is the only prior paper that
also advocates for a modular quantum chiplet design to mitigate fabrication defects. They analyze how to reduce the resource overhead in the pursuit of making a defect-free device. In contrast to~\cite{kate_chiplet}, which does not specialize in any particular application, our research specifically targets quantum error correction. This focus allows us to utilize some defective chiplets, significantly reducing the resource overhead compared to~\cite{kate_chiplet}.

Our method for adapting the surface code to defective qubit arrays builds upon earlier work on super-stabilizers~\cite{dan2017,shell_theory,shell_numerics}. However, prior studies did not establish a post-selection criterion applicable to modular chiplets, nor did they perform an analysis of the associated resource overhead.~\cite{dan2017} proposed to correct errors that occur near to defects using super-stabilizers. The idea was built from~\cite{Stace2009, Stace2010} where super-stabilizers are constructed to correct for an idealized noise model that introduces loss errors. As the authors of~\cite{dan2017} point out, it is not clear if their measurement schedule to read out super-stabilizers will give rise to a threshold.~\cite{dan2017} also introduced a protocol for adapting the surface code to arbitrary defect distributions, but the boundary deformation in the protocol only applies to the \textit{unrotated} surface code, which uses $\sim 2X$ more physical qubits per logical qubit for a commensurate code distance with the rotated code used in this paper. Our simulator includes a new algorithm for deforming boundaries due to the more complicated boundary of rotated surface code. Another difference is that we implement the shells proposed in~\cite{shell_theory} to mitigate clustered defects.

Some other methods have also been proposed to handle faulty qubits. Nagayama et al.~\cite{nagayama2017surface} also formed large stabilizers around the defects. They use SWAP gates to collect all the syndrome information onto one qubit, while we adopt the approach that takes the product of gauge operators. 
Wu et al.~\cite{wu2022synthesis} developed an algorithm to adapt surface code to devices with sparse connectivity such as the current IBM devices. According to our correspondence with them, their method is more suitable for highly symmetric lattices and is less suitable for handling arbitrary defect distributions.

We focus on static defects in this paper, but transient events such as cosmic rays could result in temporary defects. There are some strategies~\cite{suzuki2022q3de} that are specifically designed for transient defects on QEC code.~\cite{shell_numerics} recently considered producing shells around large clusters of transiently defective qubits introduced by cosmic rays. This work identified the importance of varying the schedule of super-stabilizer measurements for clustered defects.
\section{Conclusion}
Building a large device with modular chiplets provides the flexibility to throw away unwanted chiplets and arrange the rest. Such flexibility is crucial for scaling up quantum devices to support QEC in the presence of fabrication defects. In this work, we implement an automated method to adapt a rotated surface code to a defective grid and generate syndrome measurement circuits. Then, we run numerical simulations to identify effective indicators for assessing the performance of defective chiplets relative to defect-free chiplets. With these indicators, we evaluate the resource overhead of implementing an array of logical qubits with different target fidelity and under different defect rates. We also analyze how the overhead is affected by factors like the chiplet size. We found that with modularity and the super-stabilizers, the increase of resource overhead due to defects can be limited to a small factor, which is orders of magnitude better than the defect-intolerant baseline. 

We have focused on the design that allocates one patch of a surface code on each chiplet. Dividing each patch onto multiple chiplets would increase the flexibility in post-selection. However, since inter-chip links are currently $\sim 3X$ worse than on-chip links\cite{kate_chiplet}, this decision might increase the logical error rates. Whether further division of chiplets can reduce the overhead could be an interesting subject for future work.

\section{Acknowledgements}
We thank our shepherd Yufei Ding for constructive feedback throughout the ASPLOS revision process, and A. Strikis for comments on a draft of this manuscript.
This work is funded in part by EPiQC, an NSF Expedition
in Computing, under award CCF-1730449; in part
by STAQ under award NSF Phy-1818914; 
in part by the US Department of Energy Office 
of Advanced Scientific Computing Research, Accelerated 
Research for Quantum Computing Program; and in part by the 
NSF Quantum Leap Challenge Institute for Hybrid Quantum Architectures and 
Networks (NSF Award 2016136) and in part based upon work supported by the 
U.S. Department of Energy, Office of Science, National Quantum 
Information Science Research Centers. This work was completed in part with resources provided by the University of Chicago’s Research Computing Center.
FTC is Chief Scientist for Quantum Software at Infleqtion and an advisor to Quantum Circuits, Inc. BJB is grateful for the hospitality of the Center for Quantum Devices at the University of Copenhagen. 
\bibliographystyle{plain}
\bibliography{references}

\end{document}